\documentclass{article}

\usepackage{arxiv}
\usepackage{authblk}
\usepackage{subfig}
\usepackage{graphicx}
\usepackage{epstopdf, epsfig}

\usepackage[export]{adjustbox}
\usepackage{amsmath,amssymb}
\usepackage{booktabs}
\usepackage{longtable,tabularx}
\usepackage{xcolor}
\usepackage{cleveref}
\usepackage{makecell}
\usepackage{tablefootnote} % for table footnotes

\usepackage{url}

\crefformat{section}{\S#2#1#3}

\newcommand{\newpart}[1]{{\color{magenta}}}

\newcommand{\DNSsub}{\textsc{dns}}

\newcommand{\SRsub}{\textsc{sr}}
\newcommand{\SFSsub}{\textsc{sfs}}
\newcommand{\tauSFS}{\tau^\SFSsub}
\newcommand{\olDelta}{\overline{\Delta}}
\newcommand{\grad}{\nabla}

\title{Influence of adversarial training on super-resolution turbulence reconstruction} 

\author[1]{L. Nista}
\author[2]{C. D. K. Schumann}
\author[3]{M. Bode}
\author[4]{T. Grenga}
\author[5]{J. F. MacArt}
\author[6]{A. Attili}
\author[1]{H. Pitsch}

\affil[1]{Institute for Combustion Technology, RWTH Aachen University, Aachen, 52056, Germany}
\affil[2]{Department of Engineering, University of Cambridge, Cambridge, CB2 1PZ, United Kingdom}
\affil[3]{J{\"u}lich Supercomputing Centre, Forschungszentrum J{\"u}lich GmbH, J{\"u}lich, 52425, Germany}
\affil[4]{Department of Aeronautics and Astronautics, Faculty of Engineering and Physical Sciences, University of Southampton, Southampton, SO17 1BJ, United Kingdom}
\affil[5]{Aerospace and Mechanical Engineering, University of Notre Dame, Notre Dame, Indiana 46556, USA}
\affil[6]{School of Engineering, Institute for Multiscale Thermofluids, University of Edinburgh, Edinburgh, EH9 3FD, United Kingdom}

\begin{document}

\maketitle

\begin{abstract}
Supervised super-resolution deep convolutional neural networks (CNNs) have gained significant attention for their potential in reconstructing velocity and scalar fields in turbulent flows. Despite their popularity, CNNs currently lack the ability to accurately produce high-frequency and small-scale features, and tests of their generalizability to out-of-sample flows are not widespread. Generative adversarial networks (GANs), which consist of two distinct neural networks (NNs), a generator and discriminator, are a promising alternative, allowing for both semisupervised and unsupervised training. The difference in the flow fields produced by these two NN architectures has not been thoroughly investigated, and a comprehensive understanding of the discriminator's role has yet to be developed. This study assesses the effectiveness of the unsupervised adversarial training in GANs for turbulence reconstruction in forced homogeneous isotropic turbulence. GAN-based architectures are found to outperform supervised CNNs for turbulent flow reconstruction for in-sample cases. The reconstruction accuracy of both architectures diminishes for out-of-sample cases, though the GAN's discriminator network significantly improves the generator's out-of-sample robustness using either an additional unsupervised training step with large eddy simulation input fields or a dynamic selection of the most suitable upsampling factor. These enhance the generator's ability to reconstruct small-scale gradients, turbulence intermittency, and velocity-gradient probability density functions. Conversely, the supervised super-resolution CNN network lacks the capability to reconstruct these statistics. The extrapolation capability of the GAN-based model is demonstrated for out-of-sample flows at higher Reynolds numbers. Based on these findings, incorporating discriminator-based training is recommended to enhance the reconstruction capability of super-resolution CNNs. 
\end{abstract}

%\newpage
%\section{Nomenclature}

%{\renewcommand\arraystretch{1.0}
%\noindent\begin{longtable*}{@{}l @{\quad=\quad} l@{}}

%$ANN$ & Artificial Neural Network \\
%$CNN$ & Convolutional Neural Network \\
%$DNS$ & Direct Numerical Simulations \\
%$ESRGAN$ & Enhanced Super-Resolution GAN \\
%$GAN$  & Generative Adversarial Network \\
%$jPDF$  & joint Probability Density Function \\
%$HIT$ & Homogeneous Isotropic Turbulence \\
%$LES$ & Large Eddy Simulations \\
%$PDF$  & Probability Density Function \\
%$PIESRGAN$ & Physics-Informed Enhanced Super-Resolution GAN \\
%$RMSE$  & Root Mean Squared Error \\
%$SFS$   & Subgrid-Scale \\ %% $SFS$   & SubFilter-Scale \\
%$SISR$ & Single-Image Super-Resolution \\
%$TSResNet$ & SRGAN generator \\
%$TSResNet$ & PIESRGAN generator \\

%$k$ & wavenumber \\
%$U_{i}$ & velocity components \\
%$p$ & pressure \\
%$\delta_{ij}$ & Kronecker delta \\
%$\Delta$ & filter width \\
%$\eta$ & Kolmogorov scale \\
%$\nu$ & kinematic viscosity \\
%$\omega_{x}$ & vorticity along x-direction \\
%$dx$ & grid size \\
%$Re_{\lambda}$ & Reynolds number based on Taylor-microscale \\
%$dU/dx$ & normalized velocity gradient along x-direction \\
%$Lx/L_{11}$ & number of integral scales per computational box

%$\mathcal{L}_{\mathrm{gan}}$ & loss function definition of the GAN generator \\
%$\mathcal{L}_{\mathrm{gen}}$ & loss function definition of the TSResNet network 

%\end{longtable*}}

\section{Introduction}

Recent advances in experimental diagnostics and direct numerical simulation (DNS) capabilities have enabled the generation of increasingly high-fidelity turbulent flow fields. Doing so requires expensive assessments and significant computational resources due to the chaotic, multiscale nature of turbulence, making them impractical for engineering applications. Consequently, high-fidelity reconstruction of turbulent flows from limited data has been a longstanding concern. In this context, the ability to restore high-resolution (HR) from low-resolution (LR) fields is an attractive proposition. For example,  accurately representing small-scale turbulent features is necessary to capture mixing at these scales, which could enable more accurate subfilter-scale (SFS) closure models for large eddy simulations (LES)~\cite{duraisamy2019turbulence}. Similarly, in particle image velocimetry (PIV), near-wall resolution limitations constrain the ability to capture intricate flow structures close to boundaries, where high-resolution information is crucial for understanding boundary layer dynamics~\cite{10.1063/5.0023786}.

Significant strides in modern deep-learning frameworks have motivated the development of data-driven super-resolution (SR) methods for turbulent flows, which are promising compared to traditional SR techniques~\cite{fukami2023super, kim2021unsupervised}. Their objective is to design suitable neural network architectures to upsample LR ($\phi_\textsc{lr}$) fields to HR fields ($\phi_\textsc{hr}$). The resulting model should accurately invert an unknown filter operator $\mathcal{G}$, 
\begin{equation}
\phi_\textsc{hr} \approx \phi_\textsc{sr} = \mathcal{G}_{l}^{-1} * \phi_\textsc{lr} = \mathcal{G}_{l}^{-1} * \mathcal{G} * \phi_\textsc{hr} \, ,
\end{equation}
where $\mathcal{G}^{-1}_l$ is an \textit{l}$^{th}$-order approximate inverse of $\mathcal{G}$. 
The quality of the approximate inverse operator depends heavily on the choice of network architecture and training methodology and is not universal for implicit kernel functions.

Deep convolutional neural networks (CNNs) have been applied for SR, with Fukami \textit{et al.}~\cite{fukami2019super, fukami2020assessment, fukami_fukagata_taira_2021} pioneering their use for chaotic flows. They used a conventional super-resolution CNN to reconstruct turbulent velocity and vorticity fields from LR input data obtained from experimental data and numerical calculations. % compared a standard super-resolution CNN to a multi-scale model (employing skip connections with the intention of capturing multi-scale dynamics) designed to capture the small-scale structures and found the latter to be more effective in accurately reconstructing turbulent velocity and vorticity fields from extremely low-resolution input data (a sixteenth of the resolution of the DNS). %Pant \textit{et al.}~
Using a similar architecture with residual layers \cite{sandler2018mobilenetv2}, Pant \textit{et al.}~\cite{pant2020deep} trained a deep CNN on forced homogeneous isotropic turbulence (HIT) to super-resolve filtered DNS fields and addressed the tradeoff between resolution (fidelity) and computational complexity. Liu \textit{et al.}~\cite{liu} used a deep CNN model for SR reconstruction of 3D HIT, finding that training on time-series velocity fields improved the model's SR reconstruction capabilities. Recently, Zhou \textit{et al.}~\cite{ZHOU2022105382} developed a turbulence volumetric super-resolution (TVSR) model based on CNNs, trained on various Reynolds numbers, that was robust and accurate for different Reynolds numbers, though only when coupled with an approximate deconvolution method (ADM).

While SR architectures have been shown to be more accurate than analytical models for in-sample predictions (i.e., tested on data that statistically matches training data)~\cite{ZHOU2022105382, hassanaly2022adversarial}, the current understanding of SR mainly involves results obtained from supervised training of CNN models, which require labeled LR and HR data for training. The necessarily large amounts of labeled data for this, often from DNS, can be impractical to obtain. Supervised SR-CNN-based architectures do not typically generalize to out-of-sample flow conditions or domains, for they are optimized for specific data sets and objective functions, hence they remain largely untested for out-of-sample inputs. Furthermore, supervised CNN architectures do not accurately predict high-resolution details, resulting in a loss of high-wave number features and producing blurry super-resolved fields. This is a significant limitation for small-scale reconstruction requiring accurate recovery of high-frequency detail.

To address the drawbacks of the supervised CNNs training strategy for SR, generative adversarial networks (GANs) have been employed~\cite{goodfellow2020generative} that, unlike CNNs, do not rely solely on user-defined objective functions for training. Instead, GANs' super-resolving generator networks, based on CNNs, are coupled with discriminator networks that compete in semisupervised learning. In GAN-based SR, the generator produces higher-resolution versions of low-resolution fields (e.g., filtered DNS or LES), while the discriminator network attempts to differentiate between the generated high-resolution field and the real, ``ground truth'' high-resolution field  (e.g., the DNS itself). During training, the generator learns to produce samples that are indistinguishable from genuine high-resolution data. The discriminator in turn learns to judge the authenticity of the samples. As a result, both networks improve each other during the adversarial training process.

Deng \textit{et al.}~\cite{deng2019super} compared super-resolution GAN (SRGAN) \cite{ledig2017photo} and enhanced SRGAN (ESRGAN) \cite{wang2018esrgan} approaches, initially developed for image reconstruction, for the reconstruction of the flow around tandem cylinders by upscaling two-dimensional turbulence. The ESRGAN architecture outperformed the SRGAN in mean-flow metrics and fluctuation distributions. Through the use of GAN architectures with and without physical loss functions (based on mass and momentum conservation), Lee \textit{et al.}~\cite{lee2019data} studied the prediction of flow around a cylinder in laminar conditions. One should note that this approach barely differs from DNS and is thus, by default, computationally prohibitive. In reconstructing temporal data from a large time-step interval, only the physics-based loss function adequately constrained the solution space for accurate prediction of the resolved flow motion. %Subramaniam \textit{et al.}~
%\citet{subramaniam2020turbulence} proposed a 3D version of the original SRGAN to enrich low-resolution forced HIT data, recovering DNS-level quality with the addition of fine-scale features. Looking at the two-point longitudinal and transverse correlations, the GAN solution aligned very well with the DNS solution. Recently, %Bode \textit{et al.}~
Xu \textit{et al.}~\cite{10.1063/5.0149551} presented an innovative architecture that leverages transformers in conjunction with a GAN; they found this architecture to effectively reconstruct high-resolution turbulence fields for isotropic and anisotropic flows. Kim \textit{et al.}~\cite{kim2021unsupervised} addressed the challenge of reconstructing small-scale turbulence for sparsely paired LR and HR data, adopting a cycle-consistent generative adversarial network (CycleGAN). The architecture successfully reconstructs HR flow fields with DNS-quality statistics from the LES data. To drive the (unsupervised) learning process, they designed a loss function based on the assumption that the filtered-DNS (F-DNS) fields share a similar distribution with LES data. Consequently, their approach did not use unsupervised adversarial training.

%The capability to accurately approximate the hard deconvolution operation has made GAN-based networks attractive for SFS closure modeling.
Bode \textit{et al.}~\cite{bode2021using, bode2023applying} introduced a physics-informed ESRGAN (PIESRGAN) for SFS turbulence reconstruction incorporating a loss function based on the continuity-equation residual alone. Although their model was only trained on HIT data, it was able to improve scalar-mixing predictions in a reacting jet.
Nista \textit{et al.}~\cite{NISTA2022, nista2021turbulent} then investigated the generalization capability of a similar GAN architecture by evaluating the model's extrapolability to higher and lower Reynolds numbers than those used for training. They found that the ratio between the LES filter width and the Kolmogorov scale must be preserved for adequate generalization (i.e., a fixed SR upsampling window). 
Grenga \textit{et al.}~\cite{grenga_CST} investigated the ability of similar GAN models to recognize and reconstruct gradient and counter-gradient transport in low- and high-Karlovitz-number combustion regimes~\cite{MacArt2018, macart2019evolution}. In this case, only a GAN trained for both data sets adequately reconstructed the subfilter scales for both combustion regimes. 

While numerous studies have recognized the superior performance of GANs for super-resolution reconstruction over conventional supervised CNNs~\cite{kim2021unsupervised, zhang2017learning, lee2019data, subramaniam2020turbulence, nista2022SciTech}, the underlying reasons for this advantage have not been thoroughly explored. Possibilities include the flexibility in the training strategy (supervised and partially unsupervised), the use of adversarial (unsupervised) training leveraging unpaired training data, and the exploitation of generative models to model distributions, thereby improving generalization to out-of-sample data. GANs are not without drawbacks and may exhibit training instability and mode collapse, which could hinder convergence of the generator~\cite{GAN_lr}. Given these and the computational cost of training GANs, a critical evaluation of the need for SR-GANs and the influence of the discriminator on generalizability to both in- and out-of-sample flows with respect to different kernels and sizes and different flow configurations, is necessary. Moreover, the performance of GAN-based models to reconstruct spatial features and correlations, including gradients, intermittency, and probability density function tails, has not been thoroughly examined.

%\textcolor{red}{The focus of this work lies in gaining an improved understanding of the reconstruction quality achieved by GANs, specifically in accurately reconstructing fundamental turbulence properties, as spatial and structural turbulence correlations. In contrast to most of the existing literature, our analysis doesn't merely center around comparing supervised CNN and GAN-based models in the context of datasets that are statistically similar to their training input. Instead, it adds to the body of literature by assessing their performance with inputs meaningfully different from those employed during training and by analyzing the differences in the fields produced by the architectures. }
%\textcolor{red}{This work provides insights into how effectively SR architectures can extract features that are physically coherent and consistent with properties fundamental to turbulent flows from LR fields.}

This study investigates and quantifies the role of the GAN discriminator in reconstructing turbulent fields while considering the benefits and limitations of the GAN approach. To evaluate the effectiveness of adversarial training, a supervised deep-CNN model (TSResNet; Sec.~\ref{sec:TSResNet}) is directly compared to a semisupervised/unsupervised GAN (PIESRGAN; Sec.~\ref{sec:PIESRGAN}). The same CNN architecture, physics-inspired loss functions, and training data are used for the TSResNet and  PIESRGAN models. Hence, the sole difference between the two lies in the training strategy. In particular, the ability to accurately reconstruct small-scale structures of turbulence, consistent with the fundamental theory of turbulence, is investigated. Furthermore, while most studies have focused on in-sample reconstruction within a narrow range of flow conditions, we explore both models' generalizability to out-of-sample filtered fields for different filter sizes, filter kernels, and Reynolds numbers. Our aim is to develop an understanding of the consistency of the reconstructed fields with the original flow while avoiding the generation of spurious and/or unphysical fields and preserving the fundamental turbulence characteristics. Particular emphasis is placed on the improvements of these metrics brought about by adversarial training.

%%%%%%%%%%%%%%%% METHODOLOGY %%%%%%%%%%%%%%%%%%%%%

\section{Data sets and preprocessing} %Preparation data for training and testing

Three forced HIT DNS data sets at Taylor-microscale Reynolds numbers $Re_{\lambda} = 90$, $Re_{\lambda} = 130$, and $Re_{\lambda} = 350$ are considered. These are computed using the CIAO code, which uses second-order central differences, staggered meshes, and second-order implicit time advancement to solve the incompressible Navier--Stokes equations~\cite{desjardins2008high}. The calculations have mesh sizes 256 $\times$ 256 $\times$ 256 points for the low $Re$, 512 $\times$ 512 $\times$ 512 points for intermediate $Re$, and 1024 $\times$ 1024 $\times$ 1024 points for the high $Re$. The linear forcing $\textbf{f}= \textit{A} \, \textbf{u}$ proposed by Lundgren \textit{et al.}~\cite{linearForcing_lundgren} is applied, where $A$ is a forcing parameter inversely proportional to the eddy turnover time. Table~\ref{tab:simulation_parameters} reports the simulation parameters for the three data sets. SR training used only the lowest $Re$ DNS (\textit{Re90}), with the higher $Re$ DNSs (\textit{Re90} and \textit{Re90}) being reserved for testing.

After the lowest $Re$ simulation reached a statistically stationary state, 160 snapshots of the 3D velocity vector [$u_{i}$ = $(\textit{U}, \textit{V}, \textit{W})^{T}$] were extracted every five eddy-turnover times for training. To obtain low-resolution data, all available snapshots of the instantaneous velocity were filtered using explicit filter kernels of width $\Delta = 4\, dx$, corresponding to $8 \, \eta$. Our analysis is limited by memory requirements associated with the upsampling factor. As demonstrated later, the filter size considered places the cut-off wave number into the inertial range. As the primary goal is to demonstrate the effectiveness of adversarial training for SR reconstructions by comparing different training strategies, the advantage of adversarial contribution is apparent even at these upsampling factors. The training, validation, and testing data sets were composed of high-resolution (DNS) data, and the corresponding low-resolution (filtered-DNS) data. The spatial box, Gaussian, and spectrally sharp filter kernels are defined as 
\begin{align}
    \mathcal{G}_\mathrm{box}({\boldsymbol{x}})&={\begin{cases}{\frac{1}{\Delta}} \, {\qquad \text{if}}\left|{\boldsymbol{x}}\right|\leq{\frac{\Delta }{2}} \\0 \, {\qquad \text{otherwise}}\end{cases}} 
    \\
    \mathcal{G}_\mathrm{Gaussian}({\boldsymbol  {x}})&=\left({\frac{6}{\pi \Delta ^{{2}}}}\right)^{{{\frac  {1}{2}}}}\exp {\left(-{\frac  {6({\boldsymbol  {x}})^{2}}{\Delta ^{2}}}\right)},
    \\
    \mathcal{G}_\mathrm{spectral}({\boldsymbol  {x}})&={\frac  {\sin {(\pi ({\boldsymbol  {x}})/\Delta )}}{\pi ({\boldsymbol  {x}}) /\Delta }},
\end{align}
and the F-DNS fields are obtained by a discrete downsampling operation, applied independently of the filter kernel, defined as
\begin{equation}
\mathbf{f}_\mathrm{downsampling}:\, \mathbb{R}^{\Omega} \rightarrow \mathbb{R}^{\Omega/\Delta}, \, \Bar{\textbf{u}}_\textsc{f-dns} = \sum_{i=1}^{N} \sum_{j=1}^{N} \sum_{k=1}^{N} \mathcal{G} * \textbf{u}(\Delta i, \Delta j, \Delta k) \, ,
\end{equation}
where $N$ is the number of mesh points of the high-resolution field and $\Omega$ is the corresponding three-dimensional field. The latter operation ensures that the F-DNS fields have the same dimensionality as the corresponding LES fields. A comparison of filter kernels (box, Gaussian, and spectrally sharp) is given in Sec.~\ref{sec:filter}.

\begin{table}
\begin{center}
\def~{\hphantom{0}}
\begin{tabular}{ccccccccc}
\toprule
Case & \textit{N} & $Re_{\lambda}$  & \textit{$Re_t$}  & \textit{$L_x/L_{11}$} & \textit{$dx/ \eta$} & \textit{$k_{max}$}  & Train & Test \\
\midrule
Re90  & $256^3$  & 90 & 920 & 5.26 & 1.98 & 128 & $\bullet$ & $\bullet$ \\
Re130 & $512^3$  & 130 & 3320 & 5.26 & 1.86 & 256 & & $\bullet$ \\
Re350 & $1024^3$ & 350 & 16182  & 5.26 & 1.97 & 512 & & $\bullet$ \\
\bottomrule
\end{tabular}
\caption{Simulation parameters for training and testing data sets. \textit{N} is the number of mesh points, \textit{$Re_t$} is the turbulent Reynolds number, \textit{$L_x/L_{11}$} is the number of integral scales within the computational domain, \textit{$dx/ \eta$} is the mesh resolution relative to the Kolmogorov microscale \textit{$\eta$}, and \textit{$k_{max}$} is the largest wave number represented. ``Train'' and ``test'' flags indicate which data were used in model training and testing, respectively.}
\label{tab:simulation_parameters}
\end{center}
\end{table}

Loading the entire training data set is impractical given memory limitations. To alleviate this, we extract sub-domains randomly from each snapshot of the full domain. The first 120 snapshots were used for training, while the last 40 frames, equally divided, were used for validation and testing (training/validation ratio = $0.75$/$0.25$). The ongoing validation was then performed on 20 in-sample fields not used for training. The last 20 snapshots were employed to obtain averaged results. The time separation between training/validation samples and testing samples spans more than 10 large-eddy turnover times, thus the two sets of samples should be statistically uncorrelated. Both networks, TSResNet and PIESRGAN (introduced in Sec.~\ref{sec:architectures}), were trained using subboxes of the original computational domain with a size of $128 \, \eta \times 128 \, \eta \times 128 \, \eta$, corresponding to subboxes of $64 \times 64 \times 64$ and $16 \times 16 \times 16$ for high- and low-resolution fields, respectively. The size of these subboxes was found to be a good compromise between memory requirements and the inherent length scale of the turbulent flow. It is important to emphasize that the dimensions of the subboxes were constrained by both $\eta$ and the integral length scale $l_t$. Specifically, it is ensured that at least one $l_t$ is contained in these subboxes. Using those settings, 6400 subboxes were employed for the training. During the evaluation of testing samples, the entire test sample is reconstructed continuously, without employing cropped reconstruction boxes as is done for training. The velocity components of high-resolution data and low-resolution data used for training and testing were normalized by the global maximum and minimum of the DNS [i.e., $(U_\textsc{f-dns}, V_\textsc{f-dns}, W_\textsc{f-dns}) \in [0,1]$ and $(U_\DNSsub, V_\DNSsub, W_\DNSsub) \in [0,1]$] to improve the network's performance~\cite{geron2019hands}.
%I think that the important point to say is that it should be (at least) larger than the filtering order to learn some large structure. it is just the ratio of Filter size to box size, that is important.%

The large data set and deep convolutional frameworks (implemented in TensorFlow~\cite{tensorflow2015-whitepaper}), required training parallelization; this was done using the Horovod library~\cite{sergeev2018horovod}. %To circumvent memory limitations associated with batch size, the network was replicated across several workers (GPUs), splitting the training dataset among these units and updating the gradients synchronously at the end of each batch. 
Calculations were performed on the J{\"u}lich DEEP-EST cluster (DEEP-DAM partition) using four nodes, each with one NVIDIA V100 32GB GPU. This gave an overall speed-up close to a factor of four versus single-GPU training. A mini-batch of eight subboxes per GPU and the ADAM optimizer, with an initial learning rate of $10^{-4}$, were chosen due to memory constraints and previous evidence of the success of this combination~\cite{pant2020deep, lee2019data, optimizer}.
``Mixed-precision'' training is common in machine learning for its low computational cost~\cite{freytag2022impact} but can be detrimental to predictive accuracy in scientific computing.
We use single-precision arithmetic for both weight definition and loss computations, following the results of Hrycej \textit{et al.}~\cite{Hrycej_2022}. Details are provided in the Appendix \ref{app:appendix}. Moreover, during the evaluation of the testing samples, CPUs with access to a significantly larger memory pool are employed to overcome memory limitations.

For unsupervised training, LES calculations of forced HIT are conducted using the same DNS solver configuration albeit for coarser grid resolutions (equivalent to the discrete downsampling factor applied to the DNS mesh). The unresolved scales are closed using the dynamic Smagorinsky model~\cite{germano1991dynamic,lilly1992proposed} [Eq.~\ref{eqn:dynSmag}], or their influence is not explicitly modeled (``implicitly modeled'' LES). To extrapolate a given model to higher $Re$ (utilizing the \textit{Re130} and \textit{Re350} data sets for out-of-sample analyses), the filtered input field is rescaled to match the $\Delta/ \eta$ ratio used for training. This rescaling procedure guarantees a certain level of generalizability~\cite{NISTA2022}.

%%%%%%%%%% MODELS %%%%%%%%%%%%%%%%%%%%

\section{Neural network architectures and training strategy}
\label{sec:architectures}

\begin{figure}[!ht]
    \centering
    \includegraphics[width=\textwidth]{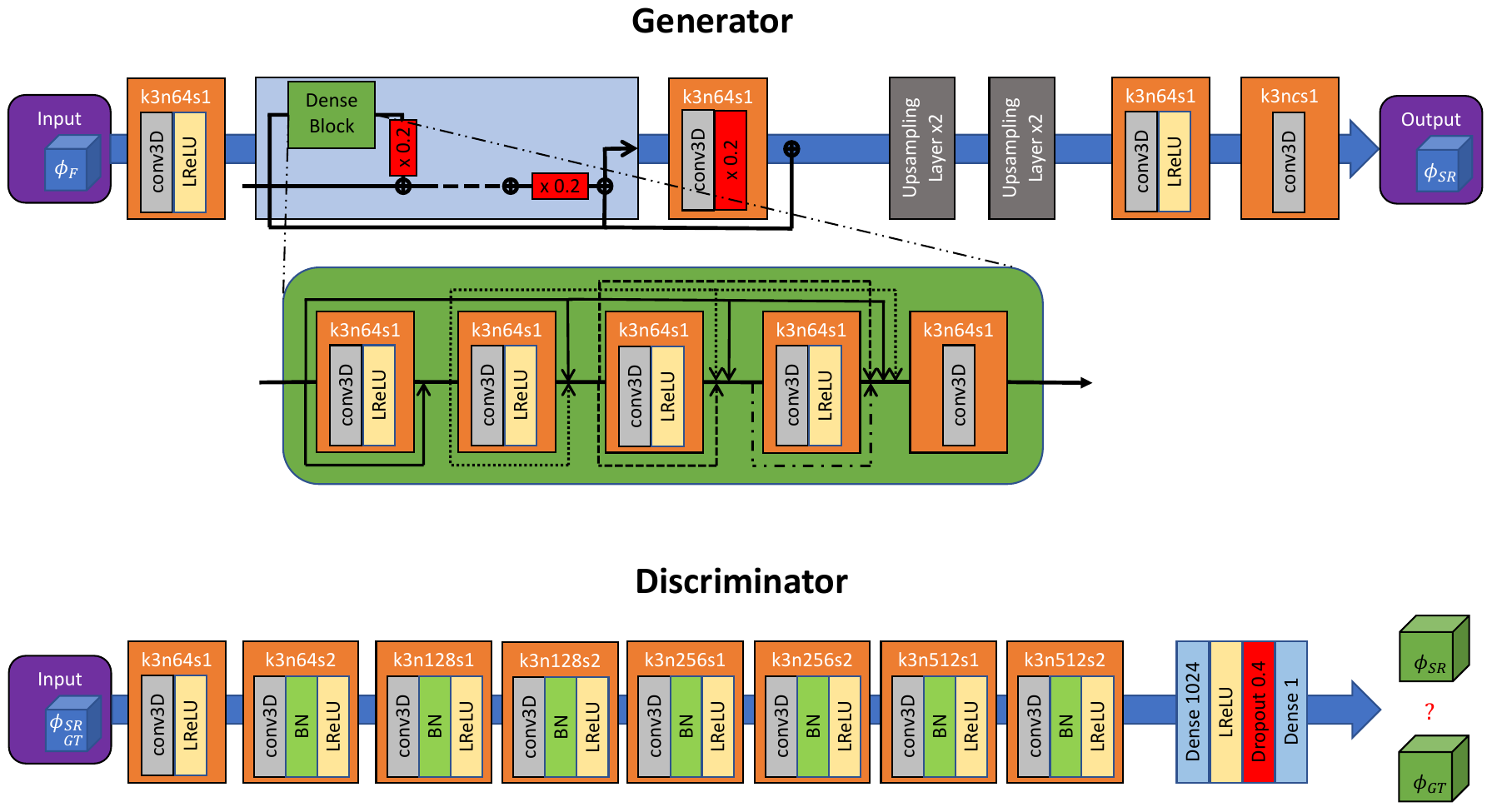}
    \caption{Top: The generator structure employed for CNN and PIESRGAN-based models. Bottom: The discriminator structure employed for PIESRGAN models.  In each, $\phi_{F}$ is the filtered input field, $\phi_{SR}$ is the super-resolved field, and $\phi_{GT}$ is the ground-truth (DNS) field. Each convolutional block contains kernels of size $k$, $n$ filter maps, and $s$ strides along each spatial dimension of the convolutional layer.}
    \label{fig:TSRGAN_architecture}
\end{figure}

Our networks are based on the architecture of Bode \textit{et al.} \cite{bode2021using, bode2023applying, bode2023ai, BODE_book}. Differing from previous work, our generator network uses additional up-/downsampling layers and dense layers, depicted in Fig.~\ref{fig:TSRGAN_architecture}; this is inspired by the original ESRGAN architecture~\cite{wang2018esrgan} and adapted for small-scale turbulence reconstruction~\cite{nista2021turbulent}. Along with the definition of the two architectures, the training approaches are described next.

\subsection{TSResNet}
\label{sec:TSResNet}

The TSResNet (Turbulent Super-Resolution Residual Network) is a supervised CNN-type model. It uses only the generator shown in Fig.~\ref{fig:TSRGAN_architecture} without the discriminator. The TSResNet is adapted from the original SRResNet \cite{he2016deep}, which can capture small-scale features when skip connections are included. The generator relies on three-dimensional convolutional layers with Leaky Rectified Linear Units (LReLUs) as activation functions, as they have higher computational efficiency compared to standard rectified linear units (ReLUs)~\cite{tensorflow2015-whitepaper}. 
The Residual-in-Residual Dense Blocks (RRDBs) include residual connections, a set of densely connected-layer blocks (three in the present work), and a residual-in-residual structure~\cite{wang2018esrgan}. Together, these enable super-resolved data to be generated through a very deep network. This capability is essential to facilitate learning complex transformations and is currently one of the state-of-the-art features of super-resolution networks~\cite{li2020survey}. Two upsampling layers increase the input spatial resolution by a factor of four in each spatial dimension. Each one doubles the dimensions of the input by nearest-neighbor interpolation (replicating adjacent grid points), followed by a convolution layer to improve the approximation. The upsampling branch (the gray section in Fig.~\ref{fig:TSRGAN_architecture}) is later modified in Sec.~\ref{sec:multiplesize}, where additional upsampling layers are added. The total number of trainable parameters in the TSResNet is approximately $18$ million.
%The main difference with the original ESRGAN structure is the lack of the upsampling layers, as the network is developed to add small-scale turbulent features without performing upsampling operations \cite{bode2021using}. 
%Hence, the input and output of the generator hold the same dimensions, but the energy distribution is enriched towards high-wavenumber frequencies. 
%The total number of trainable parameters in the TSResNet is approximately $18$ million.

The perceptual loss originally proposed by Wang \textit{et al.}~\cite{wang2018esrgan} for image reconstruction has been found to be less suitable for turbulence modeling~\cite{hassanaly2022adversarial} and is thus replaced with constraints derived from the continuity equation.
The loss function used for the TSResNet training process ($\mathcal{L}_{\mathrm{CNN}}$) is a combination of pixel loss ($L_{\mathrm{{pixel}}}$), pixel gradient loss ($L_\mathrm{{gradient}}$) and continuity loss ($L_\mathrm{{continuity}}$),
%Antonio: I just thought that one of the reasons  why all of this works could be related to the presence of a gradient (both explicitly in the beta_2 term and implicitly due to  the beta_1 and the CNN architecture). The ability to reconstruct the field might be actually related to the amplitude modulation that characterizes turbulence, see for example the following  paper I published some time ago where we showed that intense small-scale activity is related to strong large-scale gradients (Scale interactions in a mixing layer -- the role of the large-scale gradients)
\begin{equation}
\begin{aligned}
     \mathcal{L}_{\textsc{cnn}} &=  \beta_1 \, L_{\mathrm{pixel}} + \beta_2 \, L_\mathrm{{gradient}} + \beta_3 \, L_\mathrm{{continuity}} \\
     L_\mathrm{{pixel}} &= \mathrm{MSE}(\phi_\SRsub, \phi_\DNSsub) \\
     L_\mathrm{{gradient}} &= \mathrm{MSE}(\grad \phi_\SRsub, \grad \phi_\DNSsub) \\
     L_\mathrm{{continuity}} &= \mathrm{MSE}(\nabla \cdot \phi_\SRsub, 0),
\label{eqn:loss_generator}
\end{aligned}
\end{equation}
where the coefficients $\beta = [0.89, 0.06, 0.05]$ were previously selected by Bode \textit{et al.}~\cite{bode2021using} though hyperparameter tuning, $\phi_\SRsub$ is the super-resolved field, and $\phi_\DNSsub$ is the DNS field.
The mean-squared error (MSE) is computed between the reconstructed and ground-truth fields and is applied separately to all elements when tensor quantities are considered. The $L_{\mathrm{pixel}}$ loss function is inherently dimensionless, owing to the prior normalization of both the input and output fields. Likewise, the $L_{\mathrm{gradient}}$ loss function is normalized using the Kolmogorov length scale $\eta$, as the gradients are small-scale quantities. Consequently, the $L_{\mathrm{continuity}}$ function loss as the sum of normalized gradients is also dimensionless. This normalization helps to ensure that the loss functions are not affected by the choice of the grid spacing or the velocity magnitude, allowing for more general and scalable applicability of the nondimensional $\beta$ parameters across similar configurations.

The network was initialized with random weights to ensure unbiased predictions. Training occurs for a sufficient number of epochs after which the loss function and statistics computed on the reconstructed field do not change substantially. 

\subsection{PIESRGAN}
\label{sec:PIESRGAN}

In contrast, the PIESRGAN model uses the full GAN architecture, i.e.,  generator and discriminator, in adversarial training. The generator has the same architecture as the TSResNet described previously. The discriminator (Fig.~\ref{fig:TSRGAN_architecture}) is a deep deconvolutional architecture of fully connected layers with binary classification output, differing from the original ESRGAN discriminator by the introduction of a dropout layer to prevent over-fitting. The total number of trainable parameters of the discriminator is approximately $19$ million.

The generator's loss function is largely unchanged compared to the TSResNet but now includes the contribution of the discriminator loss ($L_\mathrm{{adversarial}}$~\cite{jolicoeur2018relativistic}). The generator loss function is $\mathcal{L}_{\textsc{gan}} = \beta_1 \, L_{\mathrm{pixel}} + \beta_2 \, L_\mathrm{{gradient}} + \beta_3 \, L_\mathrm{{continuity}} + \beta_4 \, L_\mathrm{{adversarial}}$, where
\begin{equation}
\begin{aligned}
L_\mathrm{{adversarial}} = - &\mathbb{E}[\log(\sigma(D(G(\phi_\SRsub)) - \mathbb{E}[D(\phi_\DNSsub)]))]\ \\
&- \mathbb{E}[\log(1 - \sigma(D(\phi_\DNSsub) - \mathbb{E}[D(G(\phi_\SRsub))]))],
\label{eqn:loss_gan}
\end{aligned}
\end{equation}
$\mathbb{E}[\cdot]$ is the expectation operator, $\sigma(\cdot)$ is the sigmoid function, and $D(\phi_\DNSsub)$ and $G(\phi_\SRsub)$ are the discriminator and generator outputs. The first term of the adversarial loss function encourages the discriminator to correctly classify HR fields as real, while the second term encourages the generator to produce SR fields that can fool the discriminator into classifying them as HR. If $\phi_\DNSsub$ is not provided during the training (unsupervised training; Sec.~\ref{sec:diff_kernel}), $L_\mathrm{{adversarial}}$ is reduced to $L_\mathrm{{adversarial}} = - \mathbb{E}[\log(\sigma(D(G(\phi_\SRsub))))]$. The weights $\beta = [0.89, 0.06, 0.05, 6 \cdot 10^{-5}]$ were chosen through hyperparameter tuning, such that the absolute value of each term in $\mathcal{L}_{\textsc{gan}}$ is of the same order. It is important to note that these $\beta$ parameters may not be universally applicable and could be case-dependent. While those parameters have proven effective for HIT configurations, their optimality may vary for different flow configurations for which further hyperparameter tuning is recommended.  The loss function, used to train the discriminator when ground truth DNS fields are available, is given by
\begin{equation}
\begin{aligned}
\mathcal{L}_{\textsc{disc}} =\ &\mathbb{E}[\log(\sigma(D(\phi_\DNSsub) - \mathbb{E}[D(G(\phi_\SRsub))]))]\ \\
&+ \mathbb{E}[\log(1 - \sigma(D(G(\phi_\SRsub)) - \mathbb{E}[D(\phi_\DNSsub)]))],
\end{aligned}
\label{eqn:disc}
\end{equation}
which is based on the relativistic average GAN loss function proposed by Jolicoeur \textit{et al.}~\cite{jolicoeur2018relativistic}.
%On one hand, while the discriminator is trained, the loss penalizes the discriminator for misclassifying a real instance as fake or a fake instance created by the generator as real. On the other hand, while the generator is trained, the loss rewards the generator if the discriminator was fooled, or it penalizes the generator otherwise.  %Inspired by Jolicoeur-Martineau's work~\cite{jolicoeur2018relativistic}, the concept of the Relativistic average GAN (RaGAN) is adopted to enhance the GAN training process. The idea is to decrease the probability that genuine data is perceived as real, contrary to the standard GAN loss where the goal is to increase the probability that generated data is perceived by the discriminator as genuine. 

The training of the GAN architecture is challenging, as the generator and discriminator networks are trained to compete against each other. %Therefore, the training algorithm has to achieve a Nash equilibrium between two opponents in the game theory \cite{holt2004nash}. 
Finding a convergence point is one of the main challenges of GAN training, and training can suffer oscillations and destabilization in the model's trainable parameters~\cite{goodfellow2020generative}. 
%%Metz et al. \cite{metz2016unrolled} showed a convergence failure in which the generator collapsed to produce only a single sample or a small family of very similar samples (mode collapse). 
%%ery often the generator and discriminator oscillate during training rather than converging to a fixed point. 
%%Vanishing gradients is another common problem for the training of GANs, in which the discriminator becomes more and more accurate such that the loss function of the discriminator drops to zero where the discriminator can easily recognize the fake (generated) data from real data \cite{wiatrak2019stabilizing}. % To cope with those drawbacks, the generator is pre-trained alone by using only the pixel loss between the generated data and the ground-truth data, i.e. imposing $\beta = [1.0, 0.0, 0.0]$ in Eqn.\ref{eqn:loss_generator}. Subsequently, the pre-trained generator will be used to initialize the GAN training. This technique was proposed by Wang et al. \cite{wang2018esrgan};
To cope with these, the generator is pre-trained in a fully supervised manner (like the TSResNet) by using the original generator loss function proposed in Eq.~(\ref{eqn:loss_generator}). Subsequently, the pre-trained generator is used to initialize the GAN training. Recent findings demonstrated that the choice of the initial learning rate for both networks is essential for the local convergence of the GAN training, and recommendations from recent literature have been adopted \cite{GAN_lr, nista2021turbulent}.
%The benefit of pre-training the discriminator has been evaluated and will be described in the next section. 
%To cope with those drawbacks, two different training methodologies are proposed:
%\begin{itemize}
%    \item the generator is pre-trained alone by using only the pixel loss between the generated data and the ground-truth data, i.e. imposing $\beta = [1.0, 0.0, 0.0]$ in Eqn.\ref{eqn:loss_generator}. Subsequently, the pre-trained generator will be used to initialize the GAN training. This technique was proposed by Wang et al. \cite{wang2018esrgan};
%    \item Similar to the previous approach, since the use of only MSE pixel loss might create samples that are overly smooth, the generator is pre-trained alone by using the original generator loss function proposed in Eqn. \ref{eqn:loss_generator}.
    %\item Some authors \cite{subramaniam2020turbulence, } proposed to pre-train also the discriminator in order to negate the advantage that the generator has because of its pre-trained weights. 
%\end{itemize}
%Moreover, as some authors reported some benefits \cite{subramaniam2020turbulence, kim2021unsupervised}, the discriminator is also pre-trained to negate the advantage that the generator has because of its pre-trained weights.

\subsection{Training strategy}

Table~\ref{tab:listofinvestigations} lists the architecture, training strategy, and training/testing fields for the specific trained models used in subsequent sections. In the supervised and semisupervised approaches, both LR and DNS field pairs are accessible, and the network utilizes this paired data set in an attempt to learn the mapping function between LR fields and to their corresponding ground-truth counterparts. This methodology is viable only when the data set contains paired data, i.e., both F-DNS and DNS data. In the unsupervised approach, only LR fields (LES data) are available without the corresponding DNS fields, thus learning relies solely on the feedback provided by the discriminator and the physics-informed loss function. The partially supervised approach is a combination of the two strategies, in which F-DNS/DNS pairs and true LES data as well as the discriminator feedback are employed for training.

The TSResNet models were trained for roughly 200 epochs, after which no additional reconstruction improvement was observed, using the \textit{Re90} data set and Eq.~(\ref{eqn:loss_generator}) as the loss function. Statistical convergence required 37 wall-time hours using four NVIDIA Tesla V100 32GB GPUs. The trained TSResNet model parameters were used to initialize the GAN generators. No GAN discriminator pre-training was applied, as this step has recently been demonstrated to have minimal influence~\cite{nista2021turbulent}. The adversarial training required an additional 11 wall-time hours using the same hardware, i.e., roughly $30\,\%$ more than the TSResNet training time, over an additional 40 epochs. To counteract the influence of the additional training time performed during the GAN training, the TSResNet models were trained for an additional 40 epochs. However, this additional training time did not impact the final reconstructed field and can be considered superfluous.

\renewcommand{\arraystretch}{0.95}
\begin{table}
\centering
\begin{tabular}{c c c c}
\toprule

Model & \thead{Training \\ strategy} & \thead{Input training -- testing fields \\ Filter kernels ($\olDelta/\Delta_\DNSsub$)} & Section \\

\midrule
\midrule

\multicolumn{4}{c}{\textit{Influence of adversarial training on in-sample predictions}} \\

\makecell{TSResNet  \\ (spectral kernel)} & Supervised & S ($4$) -- S ($4$) & Sec.~\ref{sec:insample} \\

%\vspace{0.2cm} 

\makecell{TSResNet  \\ (box kernel)} & Supervised & B ($4$) -- B ($4$) & Sec.~\ref{sec:insample} \\

%\vspace{0.2cm} 

%\makecell{PIESRGAN  \\ (spectral kernel)} & \makecell{supervised $\rightarrow$ \\ semi-supervised} & S ($4$) -- S ($4$) & \cref{sec:insample} \\
\makecell{PIESRGAN  \\ (spectral kernel)} & Semisupervised & S ($4$) -- S ($4$) & Sec.~\ref{sec:insample} \\

%\vspace{0.2cm} 

%\makecell{PIESRGAN  \\ (box kernel)} & \makecell{supervised $\rightarrow$ \\ semi-supervised} & B ($4$) -- B ($4$) & \cref{sec:insample} \\
\makecell{PIESRGAN  \\ (box kernel)} & Semisupervised & B ($4$) -- B ($4$) & Sec.~\ref{sec:insample} \\

\midrule

\multicolumn{4}{c}{\textit{Influence of adversarial training on out-of-sample filters}} \\

%\makecell{PIESRGAN  \\ (multiple filter kernel)} & \makecell{supervised $\rightarrow$ \\ semi-supervised} & B $\land$ S ($4$) -- G ($4$) & \cref{sec:filter} \\
\makecell{PIESRGAN  \\ (multiple-kernel-trained)} & Semisupervised & B $\land$ S ($4$) -- G ($4$) & Sec.~\ref{sec:filter} \\

%\vspace{0.2cm} 

%\makecell{PIESRGAN  \\ (semi-supervised appr.)} & \makecell{supervised $\rightarrow$ \\ semi-supervised $\rightarrow$ \\ semi-supervised (LES fields)} & B $\land$ S $\land$ LES ($4$) - G ($4$) & \cref{sec:filter} \\
\makecell{PIESRGAN  \\ (partially unsupervised)} & \makecell{Semisupervised and \\ unsupervised} & B $\land$ S $\land$ LES ($4$) -- G ($4$) & Sec.~\ref{sec:filter} \\

%\vspace{0.2cm} 

\makecell{TSResNet  \\ (fixed upsampling)} & \makecell{Supervised} & B ($4$) -- B ($8$) & Sec.~\ref{sec:multiplesize} \\

%\vspace{0.2cm} 

%\makecell{PIESRGAN  \\ (fixed upsampling)} & \makecell{supervised $\rightarrow$ \\ semi-supervised} & B ($4$) -- B ($8$) & \cref{sec:multiplesize} \\
\makecell{PIESRGAN  \\ (fixed upsampling)} & \makecell{Semisupervised} & B ($4$) -- B ($8$) & Sec.~\ref{sec:multiplesize} \\

%\vspace{0.2cm} 

%\makecell{PIESRGAN  ($x\#$)} & \makecell{supervised $\rightarrow$ \\ semi-supervised $\rightarrow$ \\ dynamic selection} & B ($2$ $\land$ $4$ $\land$ $8$) -- B ($\#$) & \cref{sec:multiplesize} \\
\makecell{PIESRGAN  ($x\#$)} & \makecell{Semisupervised and \\ dynamic upsampling} & B ($2$ $\land$ $4$ $\land$ $8$) -- B ($\#$) & Sec.~\ref{sec:multiplesize} \\

\midrule

\multicolumn{4}{c}{\textit{Influence of adversarial training on out-of-sample predictions (higher $Re$ numbers)}} \\

\makecell{TSResNet} & Supervised (\textit{Re90}) & B ($4$) -- B ($4$) & Sec.~\ref{sec:higher_Re} \\

%\vspace{0.2cm} 

%\makecell{PIESRGAN} & \makecell{supervised (Re90)$\rightarrow$ \\ semi-supervised (Re90)} & B ($4$) -- B ($4$) & \cref{sec:higher_Re} \\
\makecell{PIESRGAN} & \makecell{Semisupervised (\textit{Re90})} & B ($4$) -- B ($4$) & Sec.~\ref{sec:higher_Re} \\

\bottomrule
\end{tabular}
\caption{List of model configurations, training strategies, and training/testing filter kernels and sizes. The filter kernels ``B,'' ``G,'' and ``S'' abbreviate box, Gaussian, and spectrally sharp kernels, respectively. The $\land$ indicates that the combined set of the input fields is employed in the training, and the $\#$ indicates the variable upsampling factor.}
\label{tab:listofinvestigations}
\end{table}

%%%%%%%%%%%%%%%%% RESULTS %%%%%%%%%%%%%%%%%%%

\section{Effects of adversarial training on in-sample turbulence reconstruction}
\label{sec:insample}

\begin{figure}[!ht]
  \centering
  \begin{minipage}[b]{0.46\textwidth}
  \includegraphics[width=\textwidth]{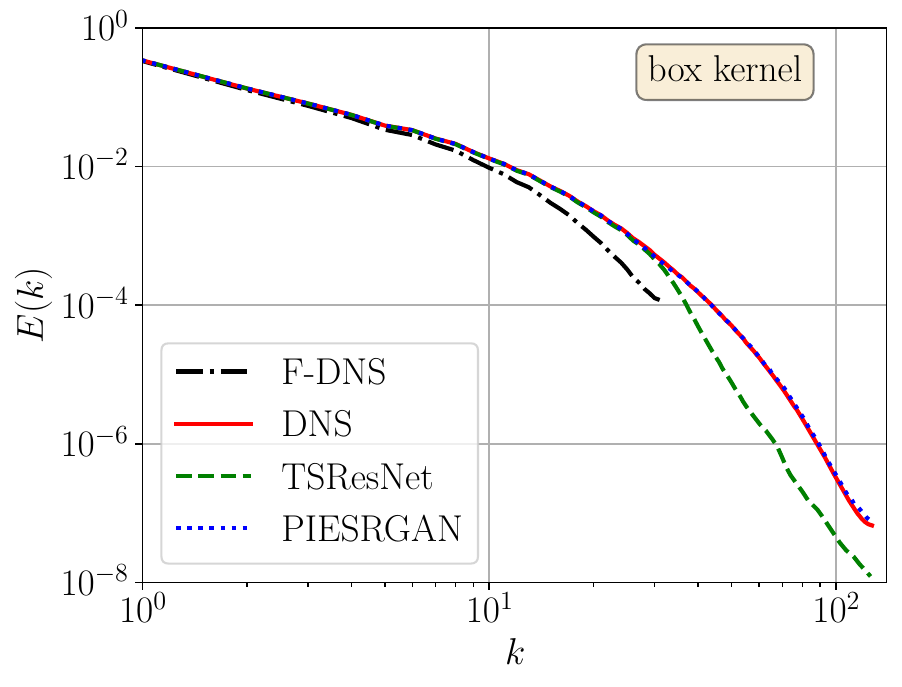}
  \end{minipage}
  \hfill
  \begin{minipage}[b]{0.46\textwidth}
    \includegraphics[width=\textwidth]{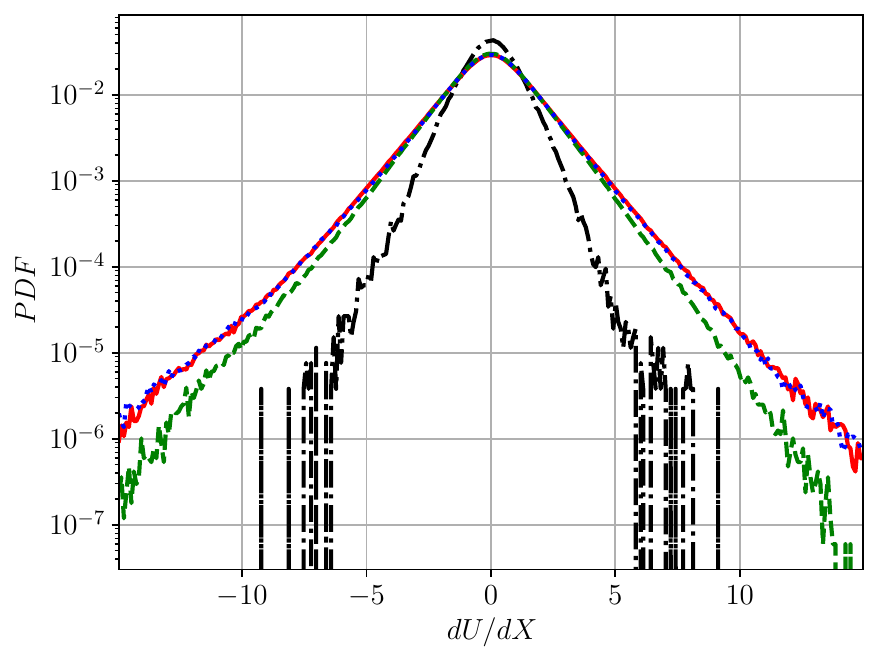}
  \end{minipage}
  \caption{In-sample (\textit{Re90}) TKE spectra (left) and PDF of the normalized velocity gradient (right) for box-filtered training/testing fields.}
  \label{fig:EnergySpectra_insample_Box}
\end{figure}

\begin{figure}[!ht]
  \centering
  \begin{minipage}[b]{0.46\textwidth}
  \includegraphics[width=\textwidth]{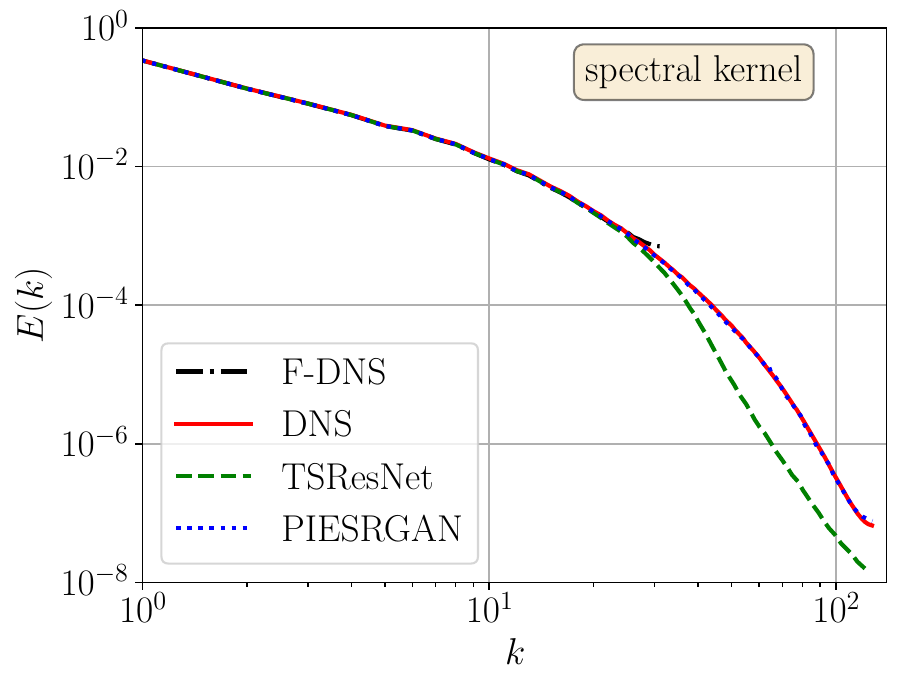}
  \end{minipage}
  \hfill
  \begin{minipage}[b]{0.46\textwidth}
    \includegraphics[width=\textwidth]{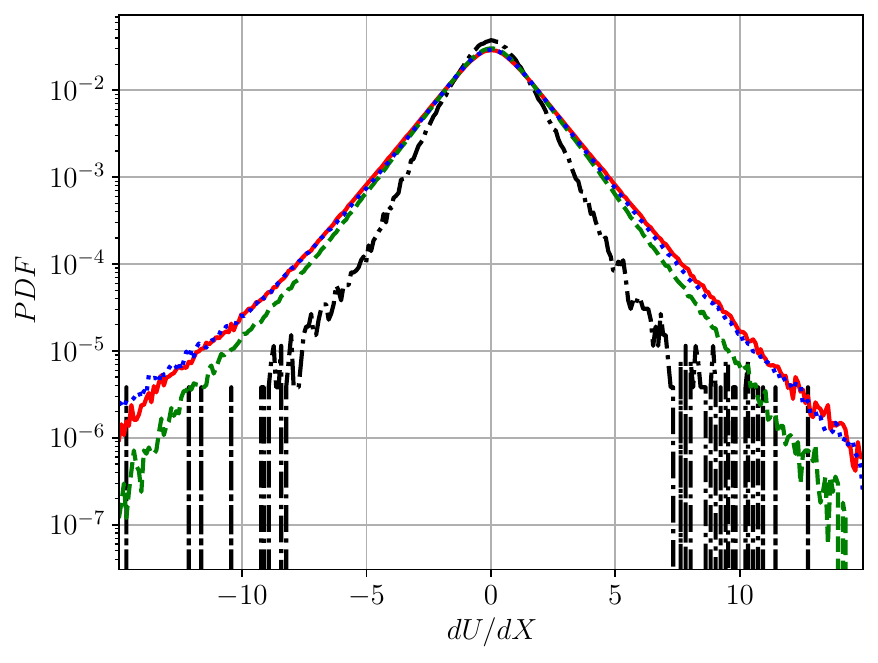}
  \end{minipage}
  \caption{In-sample (\textit{Re90}) TKE spectra (left) and PDF of the normalized velocity gradient (right) for spectrally sharp filtered training/testing fields.}
  \label{fig:EnergySpectra_insample_Spectral}
\end{figure}

Both models are tested on in-sample data, statistically matched with the training data, to assess their reconstruction capability. This evaluates the models' performance under ideal conditions.
Figures~\ref{fig:EnergySpectra_insample_Box} and~\ref{fig:EnergySpectra_insample_Spectral}  compare TKE spectra and the probability density function (PDF) of normalized velocity gradients of the filtered DNS (F-DNS), full-mesh (unfiltered) DNS, TSResNet upsampled fields, and PIESRGAN upsampled fields. The SR models are trained to produce a statistically correct SR field with DNS-like quality. The models are trained and evaluated for the \textit{Re90} case using either the box filter (Fig.~\ref{fig:EnergySpectra_insample_Box}) or spectrally sharp filter (Fig.~\ref{fig:EnergySpectra_insample_Spectral}). The primary difference in the training data is the box filter's artificial attenuation of the near-filter-scale F-DNS fields, which the spectrally sharp filter avoids.
The TSResNet recovers the TKE of the resolved scales ($k \gtrsim 32$) but fails to correctly predict the SFS energy distribution for either filter. Conversely, the PIESRGAN is more accurate, deviating from the DNS only for $k \gtrsim 122$ and energy values $E(k) \lesssim 10^{-7}$ for both filters. This behavior is quantitatively confirmed by the PDF of the normalized velocity gradient, where the TSResNet tends to underpredict the large gradients, while the PIESRGAN marginally deviates from the DNS solution. It is worth noting that the TSResNet and the PIESRGAN generator use the same network architecture, while the loss function defers only for the adversarial terms as defined in Sec.~\ref{sec:PIESRGAN}. The PIESRGAN's more accurate reconstruction is therefore due to its training, rather than the generator's network depth or loss-function differences. The PIESRGAN architecture improves the root-mean-squared error (RMSE) by $19 \%$ compared to the TSResNet for box-filtered training/testing fields and by $13 \%$ for spectrally sharp filtered training/testing fields.

\begin{figure}[!ht]
  \centering
  \begin{minipage}[b]{0.46\textwidth}
    \includegraphics[width=\textwidth]{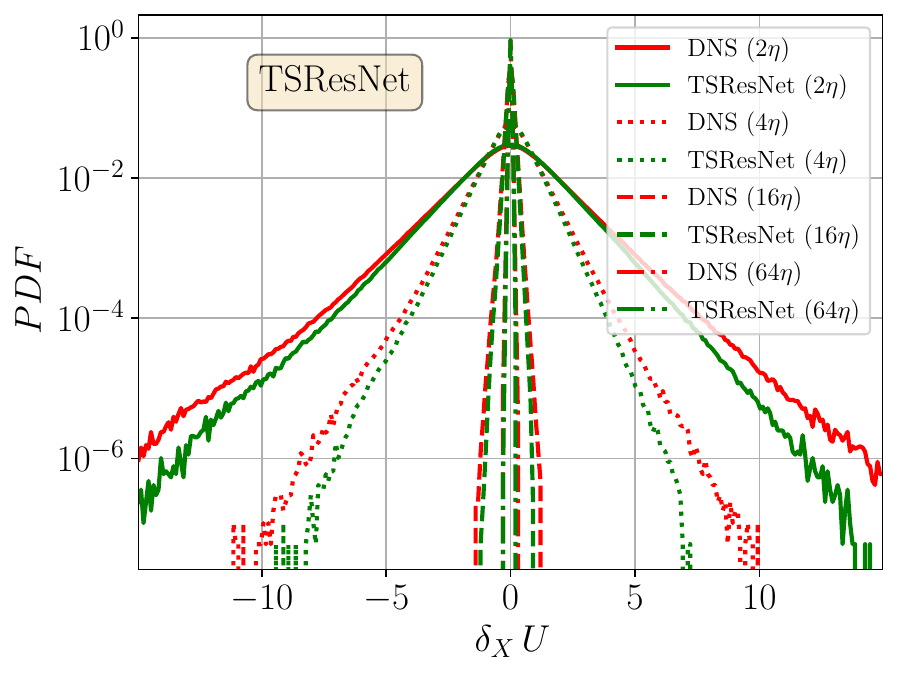}
  \end{minipage}
  \hfill
  \begin{minipage}[b]{0.46\textwidth}
    \includegraphics[width=\textwidth]{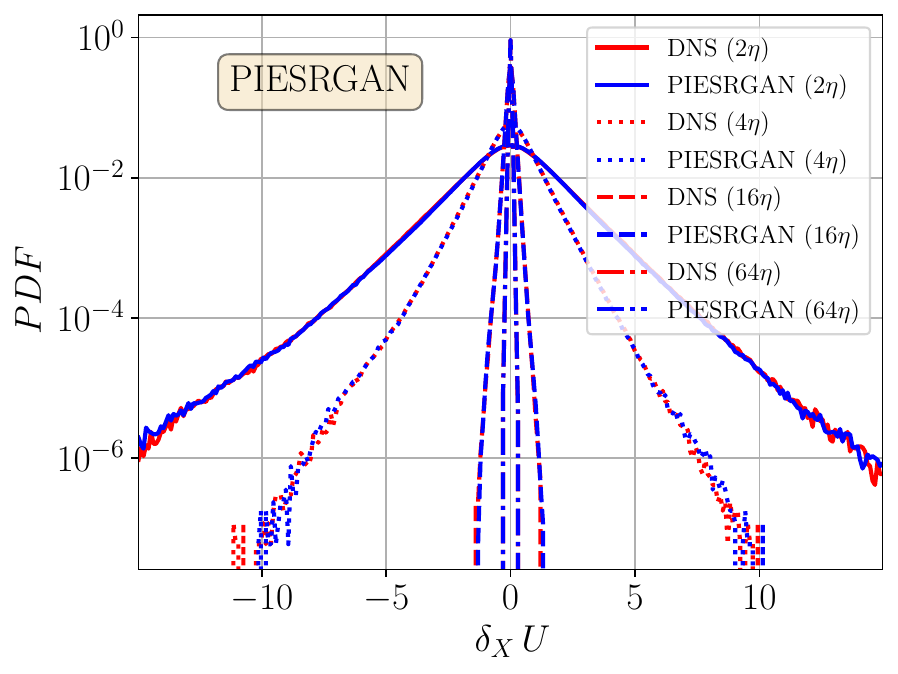}
  \end{minipage}
  \caption{PDFs of the normalized velocity increments for different separation distances for box-filtered DNS fields, TSResNet reconstructions (left), and PIESRGAN reconstructions (right).}
  \label{fig:mPDF_gradients_insample}
\end{figure}

PDFs of the normalized velocity increments $\delta_{X} U$ for different separation distances proportional to the Kolmogorov microscale $\eta$ were considered to characterize turbulent structures at different scales, similar to Attili \textit{et al.}~\cite{AntonioVelocityIncrements}.
Figure~\ref{fig:mPDF_gradients_insample} compares the DNS and reconstructed fields of velocity increments corresponding to distances of 2$\eta$ (solid lines), 4$\eta$ (dotted lines), 16$\eta$ (dashed lines), and 64$\eta$ (dash-dot lines). For small separation distances, the TSResNet captures only the most probable features well but underpredicts at the tails. For the largest increment  ($64\eta$), the gap with the ground truth is less evident. 
These TSResNet reconstructions are perceptually smoother and blurred compared to the DNS fields, as a consequence of the poor reconstruction of the large normalized velocity increments. Conversely, the PIESRGAN predictions almost exactly overlap with the DNS for all increments considered. This alignment underscores that the structures recovered by PIESGRAN are coherent with the structures that were removed during the initial filtering operation. This is due to the discriminator's nonlinear feedback to the generator (due to the discriminator's nonlinear activation functions), which boosts the generator's accuracy for the subfilter scales. This means that the GAN can reconstruct complex, nonlinear relationships between low-resolution inputs and the reconstructed fields, independent of the filter kernel applied to the HR fields. The adversarial training, therefore, improves the in-sample reconstruction capabilities compared to only using supervised learning, even with the same generator architecture.

Additionally, the SR architecture's \emph{a priori} SFS reconstruction accuracy and its effect on derived quantities are important considerations and a prerequisite for a potential \emph{a posteriori} deployment in LES. The LES-filtered momentum equation,
\begin{equation}
    \frac{\partial\overline{u}_j}{\partial t} + \overline{u}_i \frac{\partial \overline{u}_j}{\partial x_{i}} = - \frac{1}{\rho} \frac{\partial\overline{p}}{\partial x_{j}} + \nu \frac{\partial^2 \overline{u}_{j}}{\partial x_{i} \partial x_{i}} - \frac{\partial\tauSFS_{ij}}{\partial x_i},
    \label{eqn:NSmom}
\end{equation}
contains the anisotropic residual-stress tensor $\tauSFS_{ij}$, 
\begin{equation}
    \tauSFS_{ij} \equiv \overline{u_{i} u_{j}} - \overline{u}_i \, \overline{u}_j - \frac{2}{3} k^\SFSsub \delta_{ij},
    \label{eqn:tau_SFS}
\end{equation}
where $u_{i}$ are unfiltered velocity components obtained from either the DNS or the SR fields and $k^\SFSsub\equiv (\overline{u_{i} u_{i}} - \overline{u}_i\overline{u}_i)/2$ is the kinetic energy of the subfilter scales~\cite{pope2000turbulent}. In Eq.~(\ref{eqn:tau_SFS}), the $\overline{\cdot}$ operator indicates filtering (either box or spectrally sharp filters) and downsampling proportional to the upsampling factor (in the present case, the factor is equal to four). Comparisons are made to the widely used dynamic Smagorinsky model~\cite{germano1991dynamic,lilly1992proposed},
\begin{equation}
\tauSFS_{ij} = -2 c_s \overline{\Delta}^{2} \, |\overline{S}|\overline{S}_{ij}, \, \,  \mathrm{with} \, \, \,  |\overline{S}|= (2 \overline{S}_{ij} \overline{S}_{ij})^{1/2}
\label{eqn:dynSmag}
\end{equation}
where $\overline{S}_{ij}$ denotes the filtered strain-rate tensor, $\olDelta$ is the filter width, and $c_s$ is the dynamically computed Smagorinsky constant, averaged along the homogeneous directions~\cite{lilly1992proposed}. 

\begin{figure}[!ht]
\minipage{0.31\textwidth}
  \includegraphics[width=\linewidth]{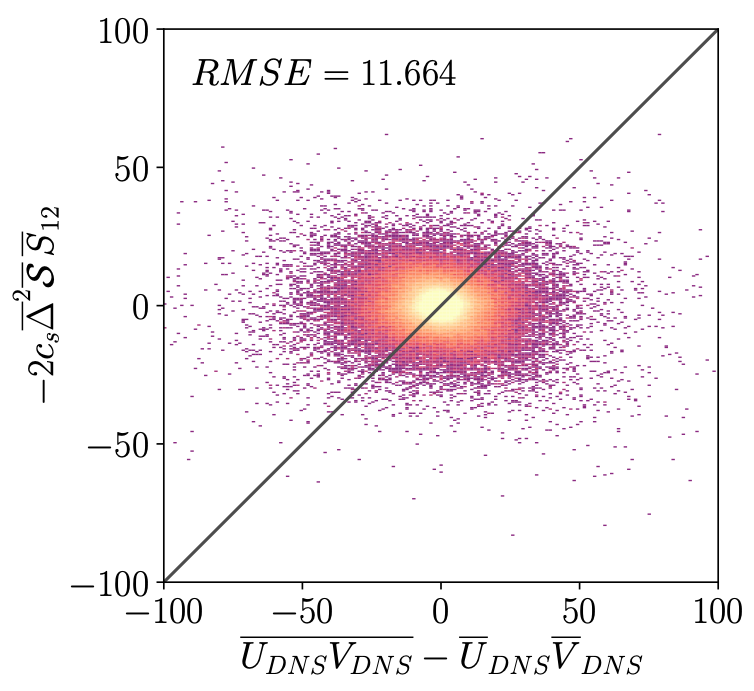}
\endminipage\hfill
\minipage{0.31\textwidth}
  \includegraphics[width=\linewidth]{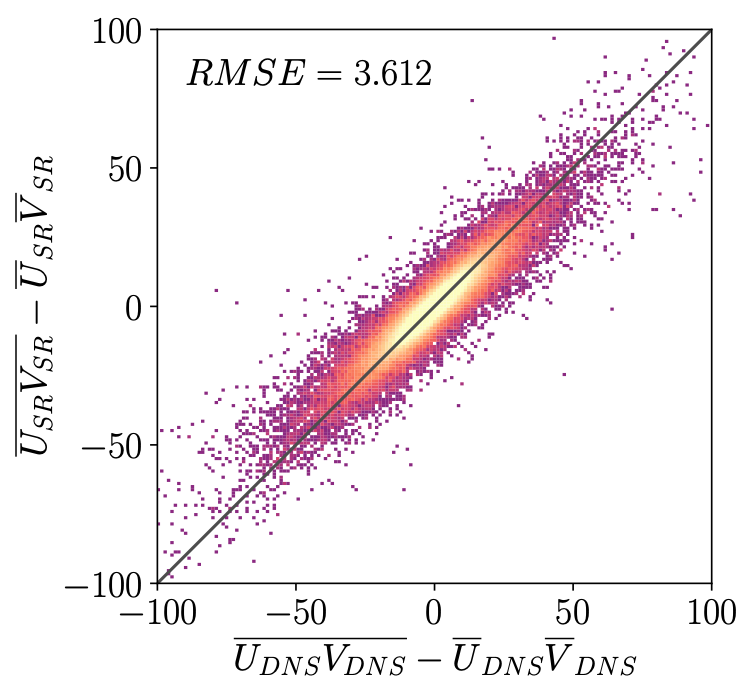}
\endminipage\hfill
\minipage{0.38\textwidth}%
  \includegraphics[width=\linewidth]{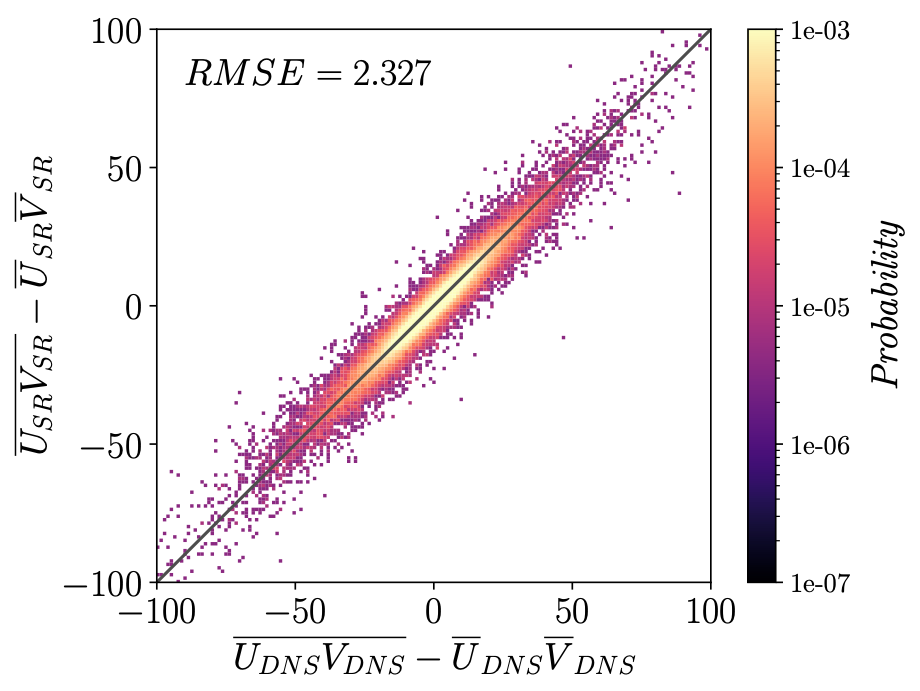}
\endminipage
\caption{Joint PDFs of the box-filtered DNS $\tauSFS_{12}$ with the dynamic Smagorinsky-modeled field (left), TSResNet reconstruction (center), and PIESRGAN reconstruction (right).}
    \label{fig:jPDF_insample}
\end{figure}

Figure~\ref{fig:jPDF_insample} shows joint PDFs (jPDFs) of the box-filtered DNS $\tauSFS_{12}$ and the dynamic-Smagorinsky (left), TSResNet-reconstructed (center), and PIESRGAN-reconstructed (right) $\tauSFS_{12}$. The PIESRGAN-reconstructed field is statistically more similar to the filtered DNS than the TSResNet-reconstructed and dynamic-Smagorinsky-modeled fields, in order of decreasing fidelity. In general, each component of the PIESRGAN-reconstructed $\tauSFS_{ij}$ has an average pointwise correlation, computed from the jPDF, with the DNS exceeding $90\,\%$. Similar results are obtained for spectrally filtered input fields. It is clear that adversarial training improves the \emph{a priori} correlation of SR fields with ground-truth data for in-sample predictions. 

\begin{figure}[!ht]
\minipage{0.31\textwidth}
  \includegraphics[width=\linewidth]{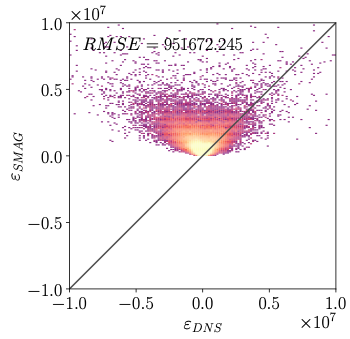}
\endminipage\hfill
\minipage{0.31\textwidth}
  \includegraphics[width=\linewidth]{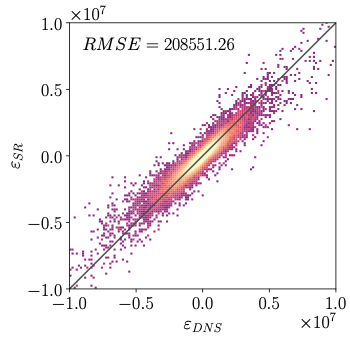}
\endminipage\hfill
\minipage{0.38\textwidth}%
  \includegraphics[width=\linewidth]{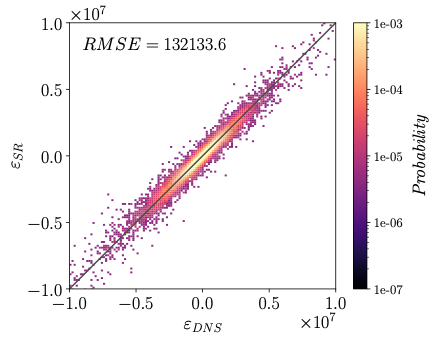}
\endminipage
\caption{Joint PDFs of the box-filtered DNS $\varepsilon^\SFSsub$ with the dynamic Smagorinsky-modeled field (left), TSResNet reconstruction (center), and PIESRGAN reconstruction (right).}
    \label{fig:jPDF_dissipation}
\end{figure}

Similarly, the SFS dissipation rate  $\varepsilon^\SFSsub = \overline{S}_{ij} \, \tauSFS_{ij}$ is highlighted because it is a crucial property of SFS models~\cite{moser2021statistical}. Figure~\ref{fig:jPDF_dissipation} shows the jPDFs of the box-filtered DNS SFS dissipation rate and those of the dynamic-Smagorinsky model (left), the TSResNet-reconstructed field (center), and the PIESRGAN-reconstructed field (right) $\varepsilon^\SFSsub$. Here, $\tauSFS_{ij}$ is determined using Eq.~(\ref{eqn:dynSmag}) for the dynamic Smagorinsky model and Eq.~(\ref{eqn:tau_SFS}) for the SR models and DNS. 

The dynamic Smagorinsky model demonstrates its dissipative nature, evident in the exclusive occurrence of positive values for $\varepsilon^\SFSsub_\textsc{smag}$. Locally, there are strong deviations from the SFS dissipation rate of the DNS. The data-driven models, on the other hand, predict dissipation rates that are in superior local agreement with the DNS. The TSResNet-reconstructed $\varepsilon_\textsc{sr}^\SFSsub$ distribution is slightly bulky around the diagonal, while the PIESRGAN-reconstructed $\varepsilon^\SFSsub_\textsc{sr}$ is closer to the correct range of the \textit{a priori} local SFS dissipation. The average SFS dissipation rate $\langle \varepsilon^{SFS} \rangle$ computed using the TSResNet is roughly $50\%$ higher compared with that computed using the PIESRGAN network. The mean SFS dissipation predicted by the PIESRGAN is within $2\%$ of the DNS value.

\newpage
\section{Accuracy of SR models for out-of-sample filters}
\label{sec:diff_kernel}
 
Recent work has highlighted practical limitations of data-driven deconvolution methods arising primarily when the application differs substantially from the training data~\cite{goodfellow2020generative,zhang2017learning, tirer2019super}. In the previous section, the PIESRGAN successfully reconstructed in-sample filtered fields $\phi_{F}$ and approximately inverted the operator $\mathcal{G}^{-1}$.
%\textcolor{red}{In the previous section, the PIESRGAN successfully  reconstructed \textit{a priori} in-sample filtered fields $\phi_{F}$ and approximately solved the full deconvolution problem. Hence, the model is essentially capable of approximating the inverse operator $\mathcal{G}^{-1}$ using analytical filtering and discrete downsampling.}
%Be clear about how the filter kernel is not invertible if a downsampling operation is also applied. Then the PIERSGAN successfully approximates the inverse filter + downsampling.
However, training SR models using only F-DNS data with corresponding full-resolution fields does not necessarily guarantee accurate upscaling of real-world fields having potentially different turbulence statistics~\cite{duraisamy2021perspectives}. The reason could be that for \emph{a posteriori} LES modeling, the interactions between grid spacing, numerical accuracy, and modeling assumptions might have to be considered. In the context of experimental analysis, the filter kernel and width implied by optical and camera systems are unknown. To understand the influence of these parameters, we now consider out-of-sample filter kernels and filter widths separately.

\subsection{Different filter kernels for training and testing} \label{sec:filter}

The trained SR models are now tested for F-DNS input fields using out-of-sample filter kernels but the same filter size. This emulates applications of trained models on fields for which the implicit filter operation is unknown, such as the use of experimental data at insufficiently high resolution. %\textcolor{red}{This emulates for example the \textit{a posteriori} application of the trained models to practical LES calculations, for which the implicit filter operation is unknown.}

\begin{figure}[!ht]
  \centering
  \begin{minipage}[b]{0.46\textwidth}
    \includegraphics[width=\textwidth]{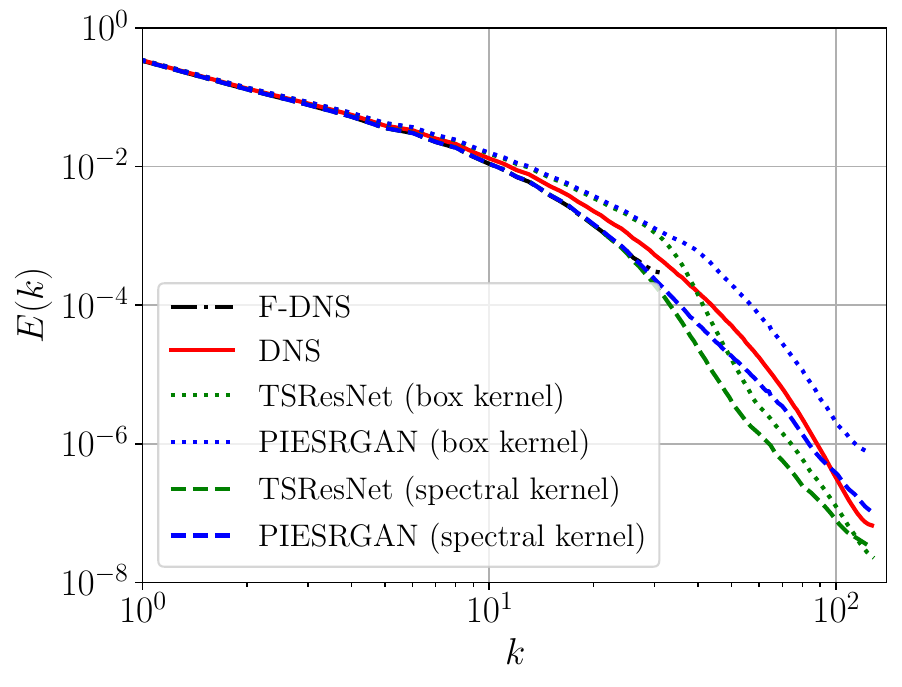}
  \end{minipage}
  \hfill
  \begin{minipage}[b]{0.46\textwidth}
    \includegraphics[width=\textwidth]{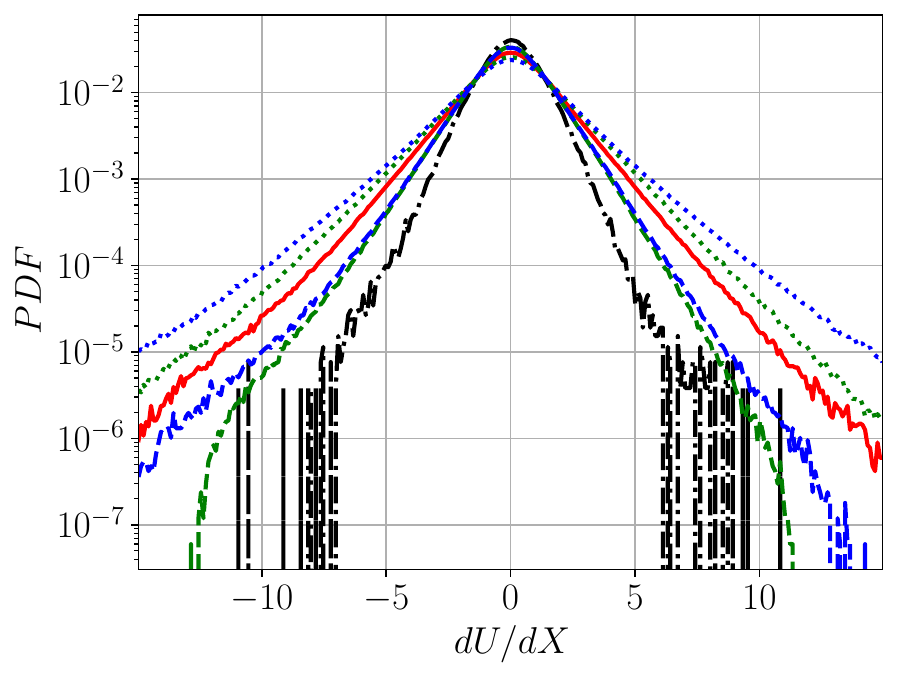}
  \end{minipage}
  \caption{TKE spectra (left) and PDF of the normalized velocity gradient (right) for models trained either for box- or spectrally sharp-filtered data and tested on Gaussian-filtered data. }
  \label{fig:filtering_inversion_bad_onlyBox}
\end{figure}

Figure~\ref{fig:filtering_inversion_bad_onlyBox} compares the TKE spectra and the PDF of normalized velocity gradients for the networks that were trained individually on box-filtered or spectrally sharp filtered fields exclusively and applied to Gaussian-filtered input data. The exclusively box-filter-trained SR models consistently overestimate the SFS kinetic energy, which results in inaccurate predictions at the high wave numbers and overpredicted energy at lower wave numbers than the filter cutoff. %This overestimation can be especially problematic in \textit{a posteriori} deployment, as it can lead to error accumulation and insufficient dissipation, resulting in unstable and inaccurate solutions. 
Conversely, for models exclusively trained on spectrally sharp filtered data, the opposite behavior is observed when the network is applied to Gaussian-filtered data: both models underpredict the resolved and SFS kinetic energy, with the PIESRGAN model being slightly more accurate than the TSResNet model. %Clearly, for unknown filter kernels (absent from the training data), as for grid-filtered LES, the performance of both SR models drops drastically.
Clearly, both SR models fail to extrapolate to filter kernels not seen in training.

One possible approach to address the reduced out-of-sample filter invertibility is to combine an SR architecture with an algebraic approximate deconvolution method~\cite{adm}. This would first use the ADM to recover the resolved velocity fluctuations, then a PIESRGAN trained for spectrally sharp filtered fields would extrapolate the subfilter scales. Still, it is important to note that the ADM-deconvoluted field would still differ from an ideal, spectrally sharp filtered field, especially near the cutoff scale~\cite{ZHOU2022105382, Zhao}. Thus, the performance of the spectral-trained GAN could suffer from its tendency to invert a known filter. A combination of an SR data-driven model and an algebraic ADM could partially minimize the GAN's over/underestimation, but it would likely not completely solve the issue.

Another potential solution leverages the GAN's flexibility to train both in a semisupervised and unsupervised mode, relying on the underlying ``structure'' of the data for patterns. This approach will be described subsequently. Both models were trained on a mixture of box-filtered and spectrally sharp filtered data in an otherwise consistent manner with the training strategy outlined previously (equal number of subboxes per filter kernel, randomly selected during the training). This approach is henceforth referred to as \textit{multiple-kernel-trained}. The discriminator is then trained to distinguish between generated high-resolution fields, upsampled from either box-filtered or spectrally sharp filtered DNS fields, and the corresponding ground-truth fields. Validating for in-sample F-DNS fields (i.e., obtained with the same filter kernels used for training), the reconstruction performance is comparable to that obtained in Sec.~\ref{sec:insample}. 
Similar to the previous analysis,  the PIESRGAN demonstrates better reconstruction capability than the TSResNet; hence we omit the TKE spectra and PDF of the normalized velocity gradient for brevity. Both models are capable of recognizing and learning both filtering operations that were applied to generate low-resolution training data, suggesting that this ability is not exclusively derived from adversarial training.

\begin{figure}[!ht]
    \centering
    \includegraphics[scale=0.70]{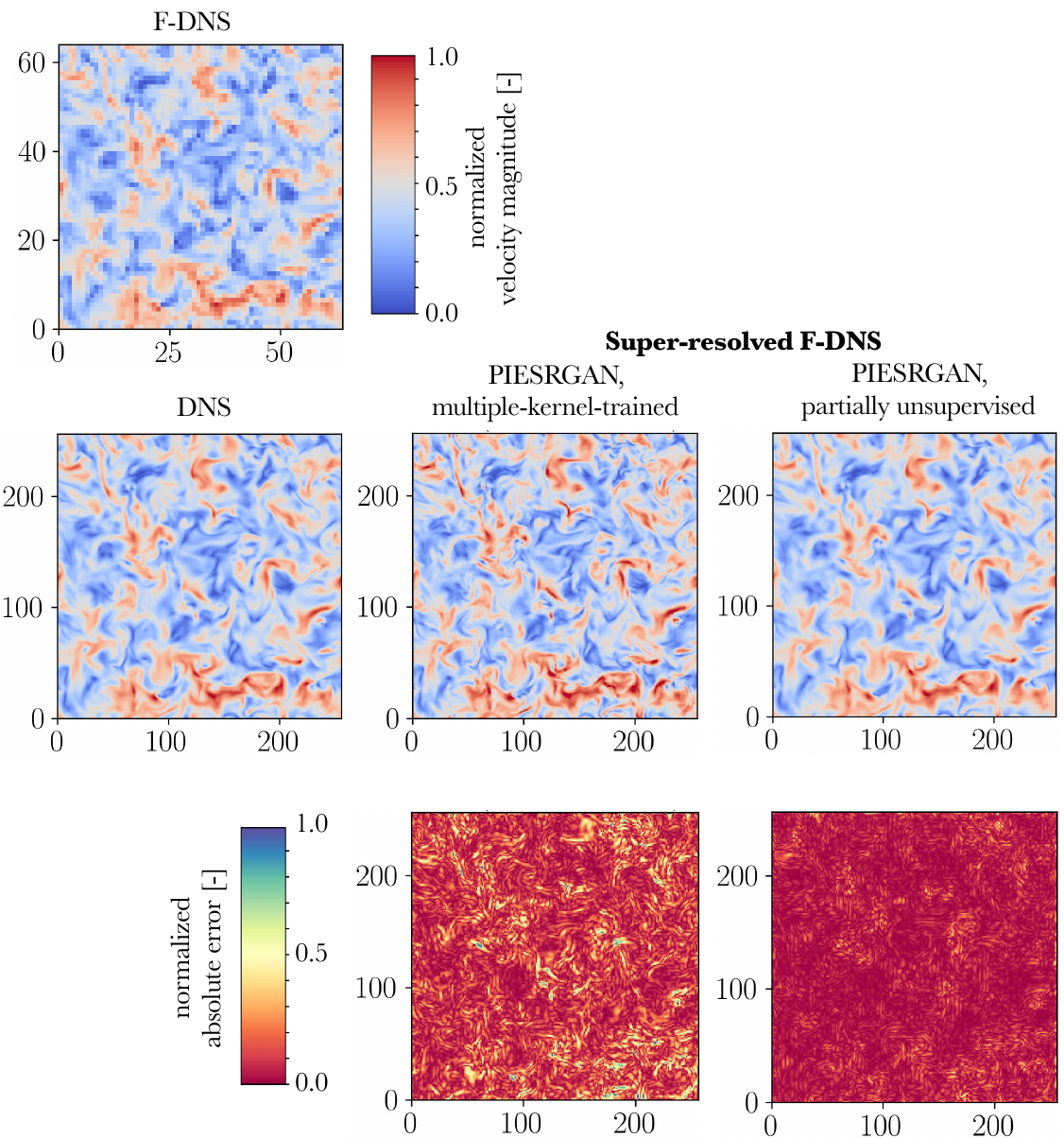}
    \caption{2D slices of instantaneous normalized velocity magnitude and the absolute error of SR fields versus DNS. The normalized absolute error is $\hat{E}=E/\max(E_{\DNSsub,\SRsub_1},E_{\DNSsub,\SRsub_2})$, where $E=\left|u_\DNSsub -{u}_{\SRsub_{1,2}}\right|$, and the subscripts $1$ and $2$ indicate the multiple-filter-kernel and semisupervised training approaches, respectively.}
    \label{fig:PlotContour_Error}
\end{figure}

\begin{figure}[!ht]
  \centering
  \begin{minipage}[b]{0.46\textwidth}
  \includegraphics[width=\textwidth]{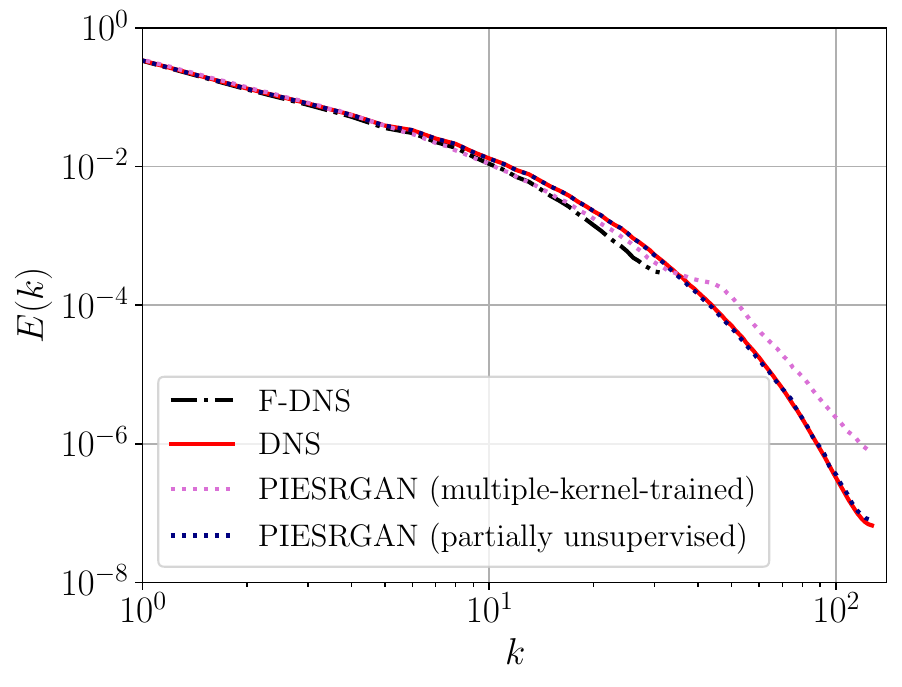}
  \end{minipage}
  \hfill
  \begin{minipage}[b]{0.46\textwidth}
  \includegraphics[width=\textwidth]{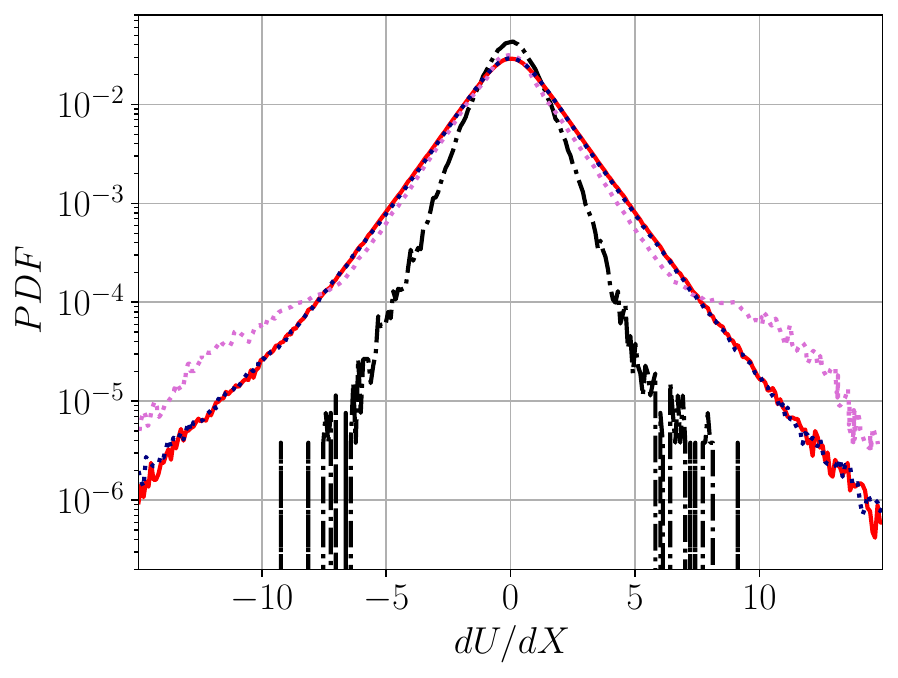}
  \end{minipage}
  \caption{TKE spectra (left) and normalized velocity gradient PDFs (right) for PIESRGAN models trained for multiple filter kernels without and with partially unsupervised step (LES-training step). Testing results are shown for Gaussian-filtered fields.}
  \label{fig:BoxSpectralLES_toGaussian}
\end{figure}

When testing for out-of-sample Gaussian-filtered input data (not included in the training data), the multiple-kernel-trained PIESRGAN model again suffers a performance drop. Figure~\ref{fig:PlotContour_Error} plots centerline slices of the normalized velocity magnitude and its absolute error obtained for the PIESRGAN using multiple-filter-kernel training. It is evident that the velocity field reconstructed by the model exhibits distortions compared to the ground-truth data. Notably, the error is especially pronounced for small-scale structures. Figure~\ref{fig:BoxSpectralLES_toGaussian} shows TKE spectra and PDFs of normalized velocity gradients for the multiple-kernel-trained PIESRGAN model, showing that the reconstructed field is in better agreement with the DNS than the single-filter-trained model, though artifacts remain in the high-frequency subfilter scales. The velocity-gradient PDF shows similar trends for the multiple-kernel-trained PIESRGAN, with consistent overprediction of the small-scale shear strain. Training for different filter kernels can mitigate filter-inversion inadequacies, but does not alleviate them altogether.
%%
%Because it is HIT, we expect all velocity gradients to behave similarly in terms of statistics right? So dU/dX is similar to dV/dX or dW/dZ, ...?
%%
%\begin{figure}
%  \centering
%  \begin{minipage}[b]{0.46\textwidth}
%    \includegraphics[width=\textwidth]{figures/DifferenzeKernels_C/Energy_onlyBoxSpectral.png}
%    %\caption{One-dimensional turbulent kinetic energy spectra when both models trained are trained with fields filtered with box and spectral filters and tested on a field filtered with a Gaussian filter.}
%    %\label{fig:shuffletraining_energy_BoxSpectral}
%  \end{minipage}
%  \hfill
%  \begin{minipage}[b]{0.46\textwidth}
%   \includegraphics[width=\textwidth]{}
%    %\caption{mPDF of the velocity gradients when both models trained are trained with fields filtered with box and spectral filters and tested on a field filtered with a Gaussian filter.}
%    %\label{fig:shuffletraining_mPDF_BoxSpectral}
%  \end{minipage}
%  \caption{One-dimensional turbulent kinetic energy spectra (left) and mPDF of the normalized velocity gradients (right) when both models trained are trained with fields filtered with box and spectral filters and tested on a field filtered with a Gaussian filter.}
%   \label{fig:shuffletraining_BoxSpectral}
%\end{figure}

While the discriminator does not directly influence the generator's ability to distinguish between filter kernels, adversarial training can still improve the accuracy of the reconstructed fields. This is an advantage of the GAN-based architecture. Inspired by Bode \textit{et al.}~\cite{bode2021using}, though differing in purpose and methodology, we now include both labeled (corresponding DNS and F-DNS fields) - as in the previous approach - and unlabeled data (computed LES fields, either with no SFS model or the dynamic Smagorinsky model, as described in Sec.~\ref{sec:architectures}) in the training process (referred to as the LES-training step). Labeled data helps the discriminator differentiate between real and fake samples, while unlabeled data improves the generator's ability to upsample realistic fields. The discriminator is trained in a supervised manner using only generated and ground-truth fields, and its weights are not updated in the LES-training step, while the generator continues to be trained on randomly sampled LES data. The resulting super-resolved LES fields are then evaluated by the trained discriminator. Since the discriminator is not trained further, the generator loss was reduced to
\begin{equation}
\mathcal{L}_{\textsc{gan}_\mathrm{unsupervised}} =  \beta_3 \, L_\mathrm{{continuity}} + \beta_4 \, L_\mathrm{{adversarial}},
\label{eqn:loss_gan_unsupervised}
\end{equation}
where $\beta_3 = 0.05$ and $\beta_4 = 6 \cdot 10^{-5}$. This is referred to as the partially unsupervised approach.

%This means that super-resolved fields generated from F-DNS and the corresponding DNS were employed to train the discriminator and the generator in the initial training stage. All four loss contributions to the loss function were active in this stage (equation \ref{eqn:loss_gan}). Meanwhile, in the second stage, the super-resolved fields generated from actual LES helped the generator architecture to generalize to multiple input fields with the help of the discriminator. The generator is thus trained to produce fields that have high-frequency details similar to an actual DNS field. This approach is similar to anomaly detection using trained discriminators. In combination with training the GAN architecture with labeled data, i.e., F-DNS (obtained with either box and spectral filters) and DNS, true LES fields produced at the same resolution of the F-DNS fields either without using an SFS model (under-resolved DNS) or with the widely-used dynamic Smagorinsky model \citep{smagorinsky1963general} were given randomly at the input. The habits of consistent overprediction at small scales is no longer present, as shown by the

The right column of Fig.~\ref{fig:PlotContour_Error} demonstrates that the PIESRGAN model, trained in a partially unsupervised manner, precisely reconstructs small-scale structures absent from the (out-of-sample) input data, with significantly lower normalized absolute error. Thus, the use of partially unsupervised training significantly reduces the overprediction of small features. In terms of RMSE, the reconstructed field from the PIESRGAN trained with multiple kernels is $67 \%$ higher compared to that computed using the partially unsupervised-trained PIESRGAN. Figure~\ref{fig:BoxSpectralLES_toGaussian} illustrates decreased overprediction of the SFS kinetic energy and better reconstruction of the normalized velocity gradient PDF when the partially unsupervised training is employed. The significant improvements at both large and small scales indicate that the unsupervised training step (LES-training step), possible only with GAN-based models, helps to generalize the model to diverse input fields.

It is worth noting that using computed LES fields or F-DNS fields as input is somehow equivalent to including random fields, as typically used in GAN applications, without incurring the risk of introducing spurious non-physical phenomena. This is important because the large scales need to be consistent between LR and HR fields. This choice enables the generator to focus on refining and augmenting existing subfilter-scale structures only, providing a guided generation process. Acknowledging the notorious instability and convergence difficulties in GAN training, we prefer this LR field-based solution over incorporating random vectors/fields.

\begin{figure}[!ht]
\minipage{0.31\textwidth}
  \includegraphics[width=\linewidth]{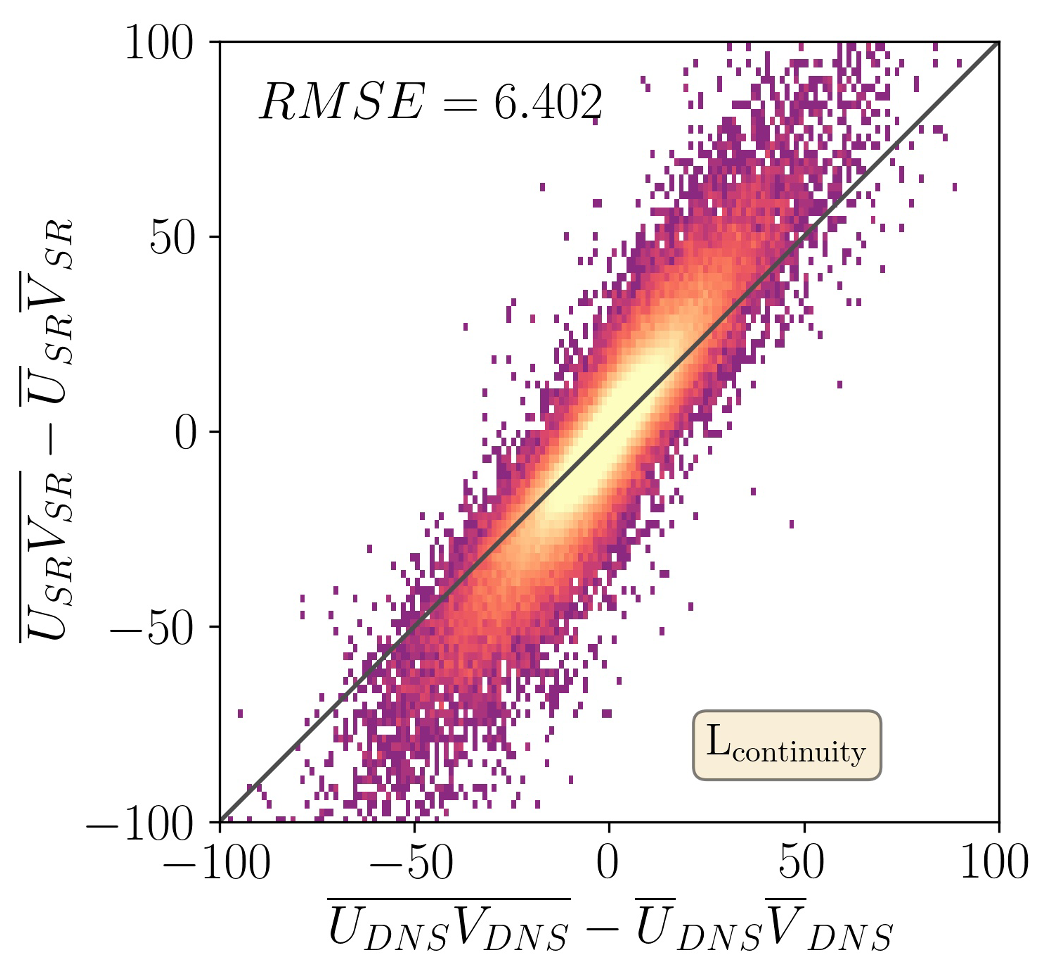}
\endminipage\hfill
\minipage{0.31\textwidth}
  \includegraphics[width=\linewidth]{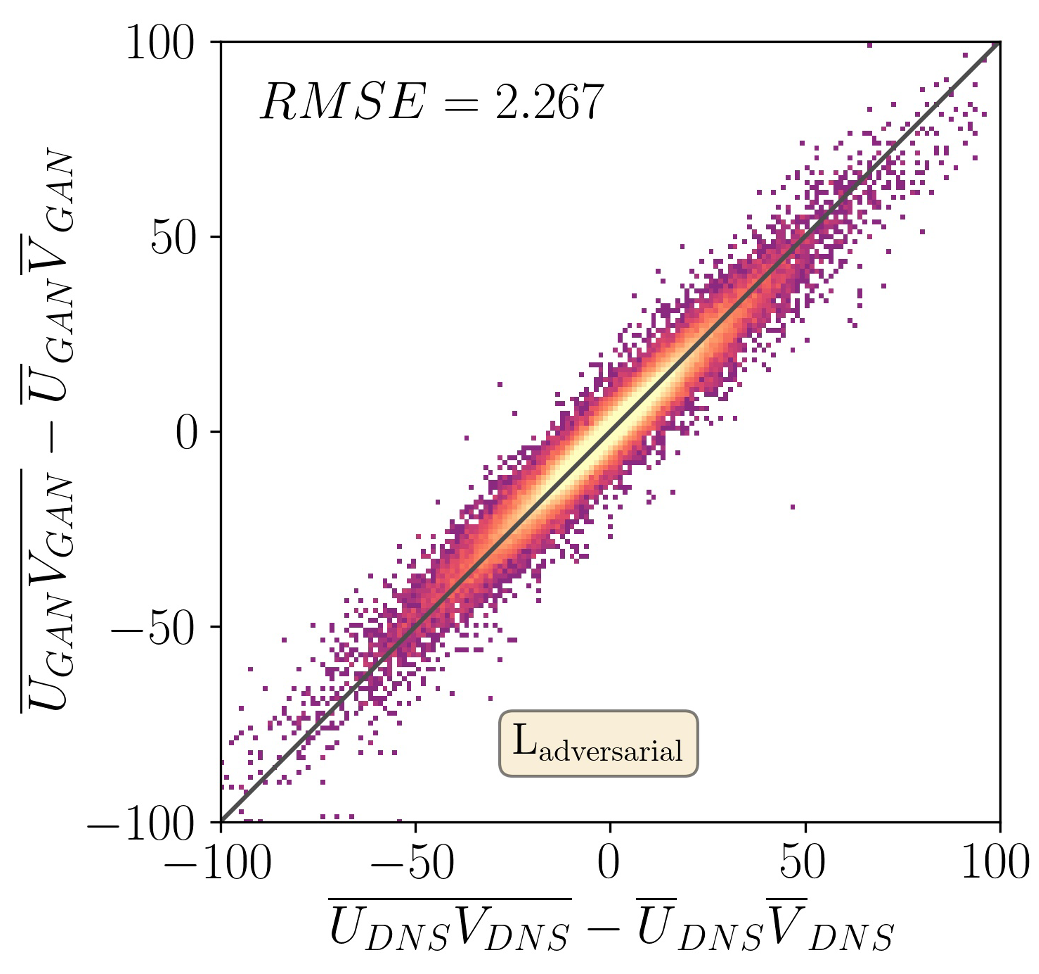}
\endminipage\hfill
\minipage{0.38\textwidth}%
  \includegraphics[width=\linewidth]{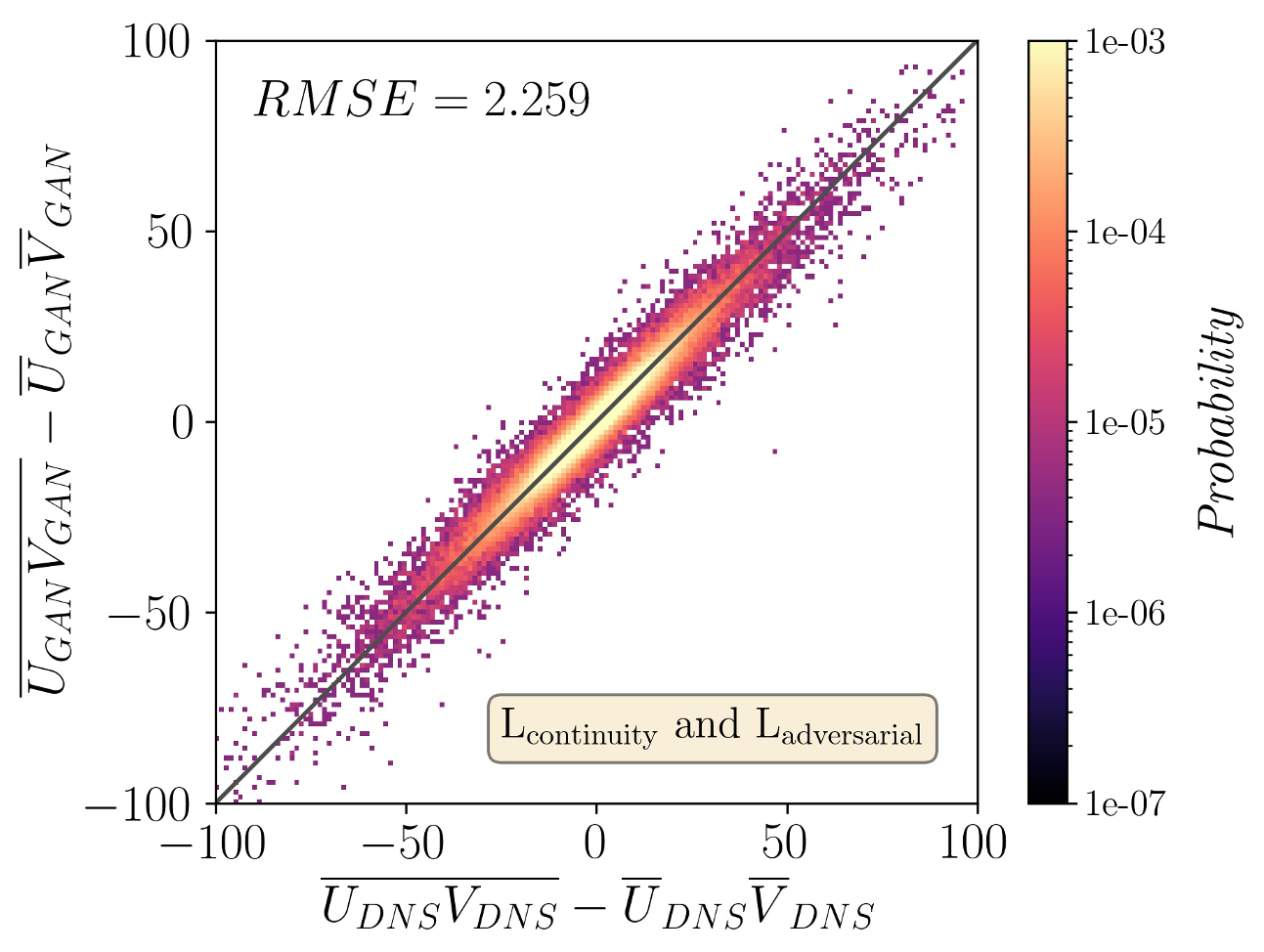}
\endminipage
\caption{Joint PDFs of the Gaussian-filtered DNS $\tauSFS_{12}$ using only $\mathrm{L_{continuity}}$ loss (left), only $\mathrm{L_{adversarial}}$ loss (center), and both $\mathrm{L_{continuity}}$ and $\mathrm{L_{adversarial}}$ losses (right) during the unsupervised training phase.}
    \label{fig:jPDF_unsupervised}
\end{figure}

To isolate the effect the discriminator exerts, we disable either the adversarial or continuity loss in Eq.~\ref{eqn:loss_gan_unsupervised}. With this, the generator is trained in a fully unsupervised manner, relying solely on feedback from the continuity equation (first option) or solely on feedback from the discriminator network (second option). Figure~\ref{fig:jPDF_unsupervised} compares the possible training options showing joint PDFs of the Gaussian-filtered DNS $\tauSFS_{12}$ [Eq.~\ref{eqn:tau_SFS}] using only $\mathrm{L_{continuity}}$ loss (left), only $\mathrm{L_{adversarial}}$ loss (center), and both $\mathrm{L_{continuity}}$ and $\mathrm{L_{adversarial}}$ losses (right) during the unsupervised training phase. The PIESRGAN-reconstructed field using only the $\mathrm{L_{adversarial}}$ loss during unsupervised training is statistically similar to that obtained when both losses are employed. Conversely, the reconstructed field obtained using only the $\mathrm{L_{continuity}}$ depicts a bulkier distribution around the diagonal and a predominant shift, resulting in a consistent overestimation of the magnitude of the Reynolds stress. While the reconstruction improvement using only $\mathrm{L_{adversarial}}$ is comparable to that shown in Fig.~\ref{fig:BoxSpectralLES_toGaussian}, the training process became more unstable and ultimately resulted in GAN collapse. To address this issue, the batch size and initial learning rate were decreased (the batch size was divided by half and the initial learning rate decreased by an order of magnitude) to improve the training stability~\cite{nista2021turbulent}. Thus, the continuity loss serves to stabilize and accelerate the training process and does not seem to meaningfully drive reconstruction performance, but further investigations are needed.
 
%Additionally, the discriminator is trained offline using only low-resolution inputs, such as actual LES data and F-DNS. 
% The second strategy is to help the discriminator to identify field content and structures at different scales. Moreover, as the low-resolution fields might already contain noise and artifacts, training the discriminator with them might make it more robust to these types of distortions. 

%%%%%%%% DIFFERENT FILTER SIZE %%%%%%%%%%%%%%
\subsection{Different filter sizes for training and testing}
\label{sec:multiplesize}

We now explore the ability of SR models to reconstruct fields for different filter sizes than from those used in training. SR training typically uses fixed upsampling factors, which can result in blurry fields and/or artifacts for upsampling factors not included in the training data. This is especially critical for flow conditions and/or geometries requiring different \emph{local} filter ratios (e.g., near-wall resolution for PIV or stretched grids for LES computations). For this, a fixed SR upsampling factor can cause over- or under-sampling, leading to reduced reconstruction capabilities. %This is especially critical in LES closure modeling, for different flow conditions and/or geometries can necessitate different \emph{local} filter ratios. For this, a fixed SR upsampling factor can cause oversampling or undersampling, leading to reduced predictive capabilities.

To investigate this, SR models trained to upsample data for an LES-to-DNS-mesh ratio $\olDelta_4 = \olDelta/\Delta_\DNSsub=4$, where $\olDelta$ is the effective LES mesh resolution and $\Delta_\DNSsub$ is the DNS mesh resolution, are tested for $\times 4$ upsampling of $\olDelta_8$ input data---that is, upsampling $\olDelta_8$ data to an effective LES resolution of $\olDelta_2$. All tests use box-filtered data. 

%both architectures trained using data filtered with a box filter with a filter size of 4 in each direction are utilized to reconstruct data filtered with a box filter with a filter size of 8 in each direction. While the PIESRGAN model can extrapolate well below the cutoff filter size, unlike TSResNet, as demonstrated by the turbulent kinetic energy spectra in 

\begin{figure}[!ht]
  \centering
  \begin{minipage}[b]{0.46\textwidth}
    \includegraphics[width=\textwidth]{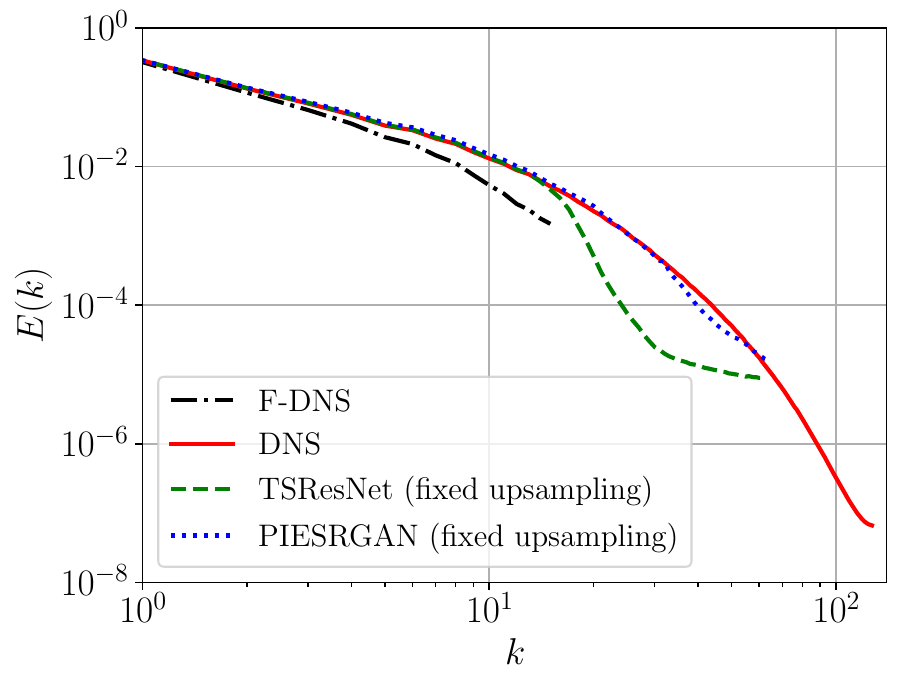}
  \end{minipage}
  \hfill
  \begin{minipage}[b]{0.46\textwidth}
  \includegraphics[width=\textwidth]{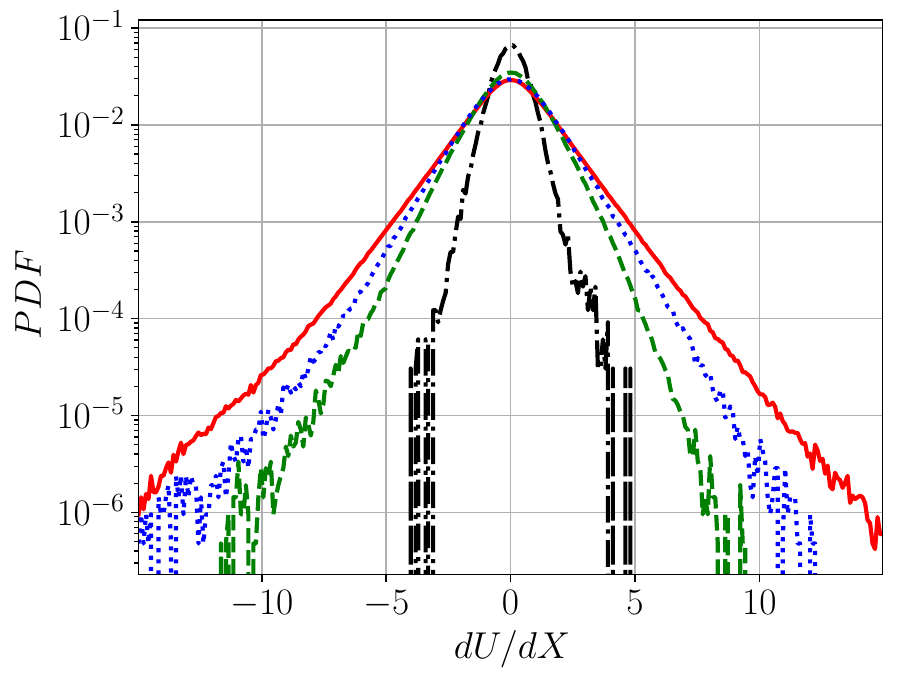}
  \end{minipage}
  \caption{TKE spectra (left) and the PDF of the normalized velocity gradients (right) when both models are trained with fields filtered with a box filter with a size of four and tested on a field filtered with a box filter with a size of eight.}
  \label{fig:differentsize_x4_Tox8}
\end{figure}

Figure~\ref{fig:differentsize_x4_Tox8} shows TKE spectra and normalized velocity-gradient PDFs for the previously supervised box-trained TSResNet and PIESRGAN models applied to box-coarser \textit{Re90} input data. The PIESRGAN accurately recovers the TKE spectrum over the upsampled range, even though it does not fully recover the low-probability tails of the velocity-gradient PDF. The TSResNet model does not fare as well, with significant deviations from the DNS spectrum throughout the upsampled range.
Despite these imperfections, it is clear that the PIESRGAN exhibits an improved tendency to successfully upsample across different scale ranges than those used for training, at least within the self-similar scales of statistically stationary HIT. Of course, this does not imply equal success for anisotropic flows.

%that using an upsampling factor of the network smaller than the optimal, where optimal refers to an upsampling factor consistent between training and testing, can be useful when the flow features and dynamics exhibit scale-similarity behavior, implying that similar patterns and structures are repeated at different scales. In such cases, a small upsampling factor could capture the important flow features without introducing significant errors and can provide a computationally efficient solution. On the other hand, using a larger upsampling factor than the optimal, the model might exacerbate noise and other uncertainties in the data. These results are presented later.

\begin{figure}[!ht]
  \centering
  \begin{minipage}[]{0.48\textwidth}
  \vspace{0pt}
  \includegraphics[scale=0.38]{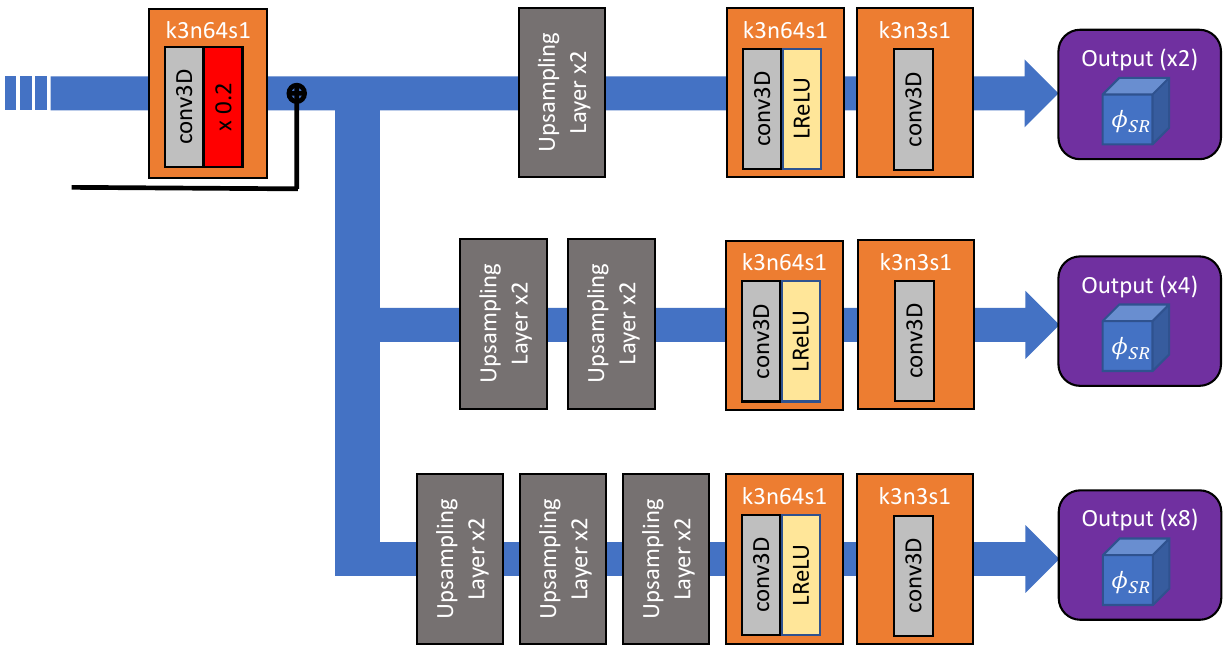}
  \end{minipage}
  \hfill
  \begin{minipage}[]{0.45\textwidth}
  \vspace{0pt}
    \includegraphics[width=\textwidth]{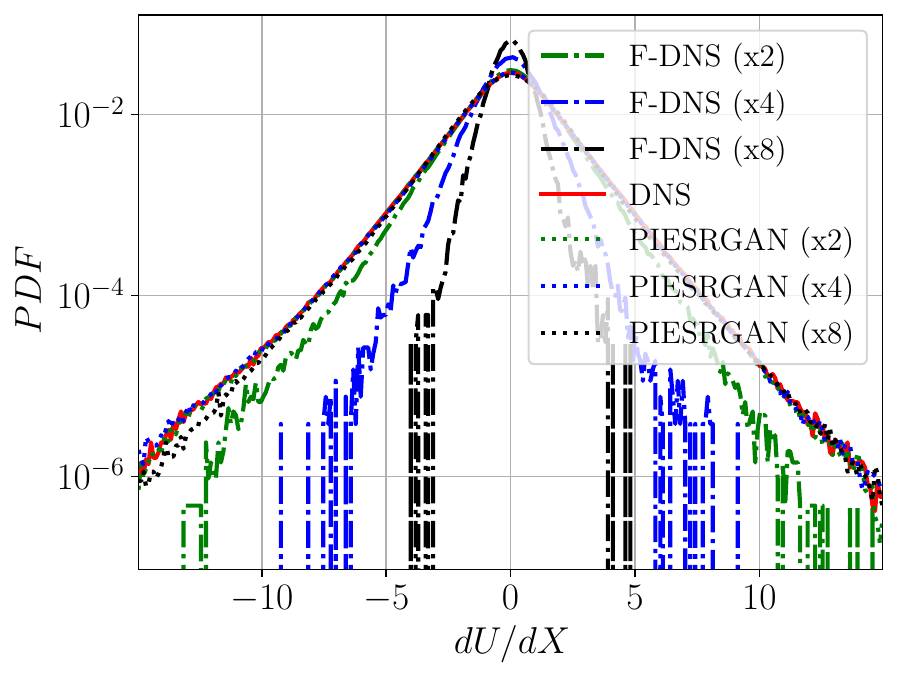}
  \end{minipage}
  \caption{\textit{Left}: Sketch of the three upsampling branches in the modified GAN generator. \textit{Right}: PDF of the normalized velocity gradient for PIESRGAN fields reconstructed using the most appropriate upsampling factor selected by the discriminator.}
  \label{fig:upsamplingLayer_and_mPDF_differentsize_together}
\end{figure}

Multiple upsampling branches can be incorporated into the generator network to determine the ideal upsampling ratio from the input data, as applied by Lim \textit{et al.}~\cite{lim2017enhanced} for image-processing tasks. We now modify the generator architecture (Fig.~\ref{fig:TSRGAN_architecture}) for multiple upsampling branches, shown in Fig.~\ref{fig:upsamplingLayer_and_mPDF_differentsize_together}. Three branches enabling different upsampling ratios are present.
\begin{enumerate}
    \item The first branch, for high-resolution inputs, contains one $\times 2$ upsampling layer.
    \item The second branch, for intermediate-resolution inputs (used in Sec~\ref{sec:insample}), contains two upsampling layers for an overall $\times 4$ upsampling ratio.
    \item The third branch, intended for low-resolution inputs, contains three upsampling layers for an overall $\times 8$ upsampling ratio.
\end{enumerate}

%The first branch is intended for input fields of very low resolution, corresponding to a factor of 8 in each direction between the HR and LR field, and consists of three upsampling layers with a size of 2 each. The second branch, used in Sec.~\ref{sec:insample}, is optimized for intermediate resolution inputs, where there exists a factor of 4 in each direction between the HR and LR field, and consists of two upsampling layers with a size of 2 each. The third branch is optimized for high-resolution inputs, with a factor of 2 in each direction between the HR and LR field, and consists of only one upsampling layer with a size of 2 in each direction (figure \ref{fig:TSRGAN_architecture}). Therefore, the current structure provides three distinct upsampling factors in the generator, referred to as the \textit{x2}, \textit{x4}, and \textit{x8} branches. 
%However, the network is only capable of handling input sizes that are known in advance during training, which limits its capability to extrapolate to various flow conditions, as the input size is not a reference parameter. A suitable branch (\textit{x2}, \textit{x4}, or \textit{x8}) should instead be determined dynamically based on the input field size. 

Multiple branch employment is promising for image processing \cite{lim2017enhanced}, though the generator needs prior information on the input size, resulting in the network being able to handle only pre determined, known input sizes. Small upsampling factors typically result in adequate reconstruction, if incompletely resolved, with larger structures typically accurately captured. Conversely, over-upsampling can lead to significant spurious artifacts \cite{NIPS2014_1c1d4df5}.

Instead, we use the discriminator to select the appropriate upsampling branch. The GAN is trained for box-filtered fields of different filter sizes ($\olDelta_2$, $\olDelta_4$, and $\olDelta_8$). During training, the correct upsampling branch is enabled based on the known filter size, and unused branches are deactivated. %This allows the network to be trained on different filter sizes, while the discriminator is trained using both the super-resolved output and the DNS field, as done before. 
During testing, the generator upsamples the input field along all three branches, producing $\times 2$, $\times 4$, and $\times 8$ fields, which the trained discriminator then uses to calculate an adversarial loss [using Eq.~(\ref{eqn:disc}) with $\phi_\DNSsub=0$]. The upsampling factor corresponding to the lowest adversarial loss is chosen. Thus, the discriminator learns to recognize high-quality fields during training and can provide highly nonlinear feedback to the generator on the quality of the generated field. 

Table \ref{tab:differentupsampling} lists a ``quality score''
\begin{equation}
    Q = \mathbb{E}[\log(\sigma(D(G(\phi_\SRsub))))],
\end{equation}
related to the discriminator loss, for $\olDelta_2$, $\olDelta_4$, $\olDelta_8$, and $\olDelta_{16}$ input data upsampled along the $\times 2$, $\times 4$, and $\times 8$ branches. Higher values indicate higher confidence (by the discriminator) in the quality of the upsampled field. The $\olDelta_{16}$ input data are out-of-sample. The branch providing the highest quality score is selected as the most-appropriate output for arbitrary input data.
\begin{table}
  \centering
  \begin{tabular}{ c > {\centering}p{1.45cm}  p{1.cm} p{1.cm}  }
    \toprule
    & \multicolumn{3}{c}{Upsampling factor} \\
    Input & $\times 2$ & $\times 4$ & $\times 8$\\
    \midrule
    $\olDelta_2$  & 0.93 & 0.46 & 0.08  \\
    $\olDelta_4$ & 0.73  & 0.91 & 0.23 \\
    $\olDelta_8$ & 0.53  & 0.69 & 0.84  \\
    $\olDelta_{16}$ & 0.17  & 0.34 & 0.59 \\
    \bottomrule
  \end{tabular}
  \caption{Quality score $Q$ for different input filter-to-DNS-grid ratios and upsampling branches.}
  \label{tab:differentupsampling}
\end{table}

Figure~\ref{fig:upsamplingLayer_and_mPDF_differentsize_together} shows normalized velocity-gradient PDFs for PIESRGAN-upsampled fields using the discriminator-selected upsampling branches. It is evident that the discriminator accurately selects the appropriate branch, enabling the modified PIESRGAN to accurately recover both the large scales and the tails of the PDFs. This adaptive upsampling strategy, utilizing the discriminator, is particularly effective in scenarios where the input filter size is not present in the training data. For example, in Table \ref{tab:differentupsampling}, the $\times 8$ upsampling branch is correctly selected as the most appropriate available branch for the out-of-sample $\olDelta_{16}$ input data.
However, it is important to note that this approach entails higher computational cost compared to using a single upsampling factor (roughly 50\,\% higher), as more fields are employed.%Additionally, if the optimal upsampling factor is not included in the generator architecture, the performance may be suboptimal, as evidenced by the last bottom row of table \ref{tab:differentupsampling}. %In conclusion, this dynamic upsampling feedback can only be achieved with a GAN-based architecture, as the discriminator plays a critical role in evaluating the reconstructed field produced by the generator.
In conclusion, this dynamic upsampling feedback, which can be provided only in the context of a GAN-based architecture, is an effective approach to mitigating the fixed upsampling/filter size limitation.

%%%%%%%%%%%%%%%% EXTRAPOLATION AT HIGHER RE %%%%%%%%%%%%%%%%%%%%%%%%%%
\section{Extrapolation to higher Reynolds numbers}
\label{sec:higher_Re}

%The previous sections demonstrate that the PIESRGAN architecture can reconstruct unresolved fields more accurately than the TSResNet for in-sample flows. Moreover, the discriminator, hence adversarial training, permits greater flexibility to moderately out-of-sample input data (e.g., for different filter kernels and/or upsampling ratios), though the model must also generalize well beyond its training flows to perform accurately in \emph{a posteriori} LES.
The previous sections demonstrate that the PIESRGAN architecture can reconstruct unresolved fields more accurately than the TSResNet for in-sample flows. The discriminator, playing a crucial role in adversarial training, permits greater flexibility for out-of-sample conditions (e.g., for different filter kernels and/or upsampling ratios), though the model must also generalize well beyond the flow conditions used for training.

\begin{figure}[!ht]
  \centering
  \begin{minipage}[b]{0.46\textwidth}
    \includegraphics[width=\textwidth]{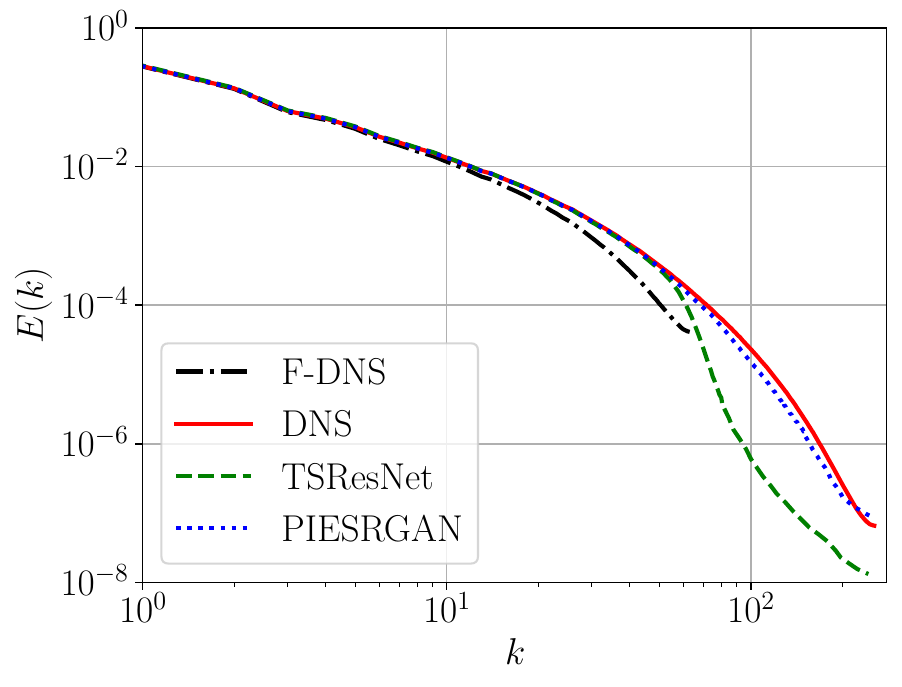}
  \end{minipage}
  \hfill
  \begin{minipage}[b]{0.46\textwidth}
    \includegraphics[width=\textwidth]{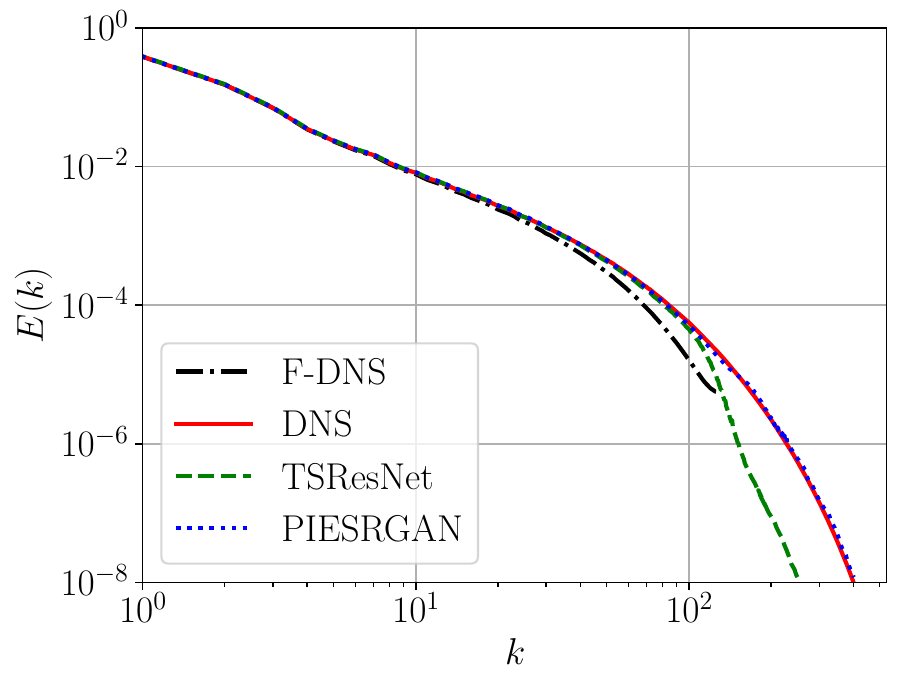}
  \end{minipage}
  \caption{TKE spectra for \textit{Re90}-trained models evaluated on higher-Reynolds-number \textit{Re130} (left) and \textit{Re350} (right) input data. Filtered DNS data were obtained using a $\olDelta_4$ box filter.}
  \label{fig:extrapolation_highRe}
\end{figure}

To test this generalization capability with equal filter kernel and size for both training and testing, thus under ideal conditions, the models trained for the lower $Re_\lambda=90$ case (\textit{Re90}) (Sec.~\ref{sec:insample}) are applied to forced HIT for the higher $Re_\lambda=130$ (\textit{Re130}) and $Re_\lambda=350$ (\textit{Re350}) cases (Table \ref{tab:simulation_parameters}). Figure~\ref{fig:extrapolation_highRe} compares the TKE spectra for these out-of-sample model applications to the unfiltered and box-filtered ($\olDelta_4$) DNS spectra.
The $Re_\lambda=90$-trained TSResNet model fails to correctly reconstruct the subfilter scales, similar to its performance in the previous in-sample analyses, while the $Re_\lambda=90$-trained PIESRGAN model accurately extrapolates the subfilter scales at these higher Reynolds numbers, at least outside the dissipation range. The ability of the PIESRGAN model to extrapolate the small scales of anisotropic flows remains to be studied.
%%%% IDEA CHRISTOPH
%train the CNN on a higher Re and compare it to the GAN trained at lower Re
%%%
 
\begin{figure}[!ht]
\minipage{0.46\textwidth}
  \includegraphics[width=\linewidth]{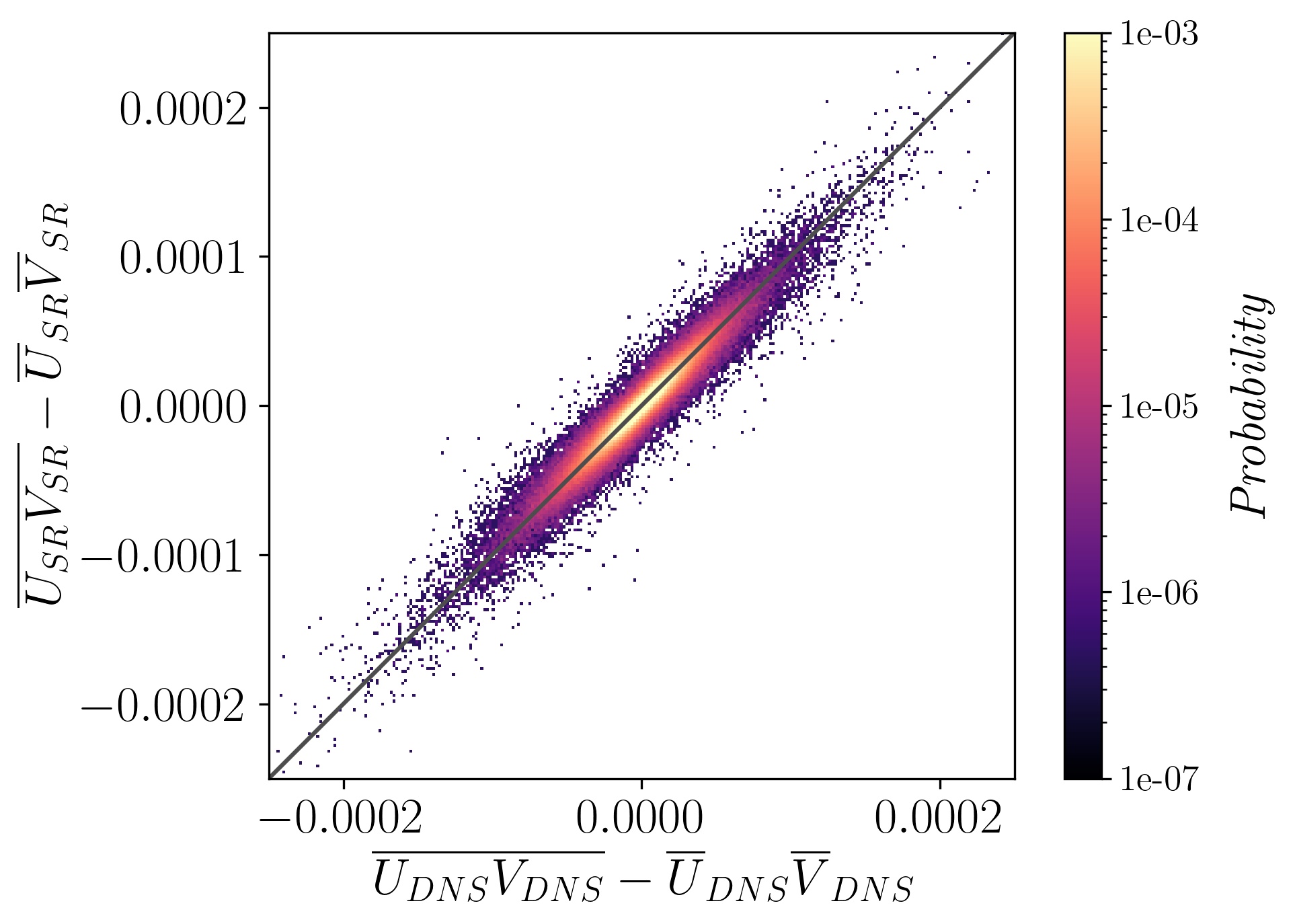}
  
\endminipage\hfill
\minipage{0.46\textwidth}
  \includegraphics[width=\linewidth]{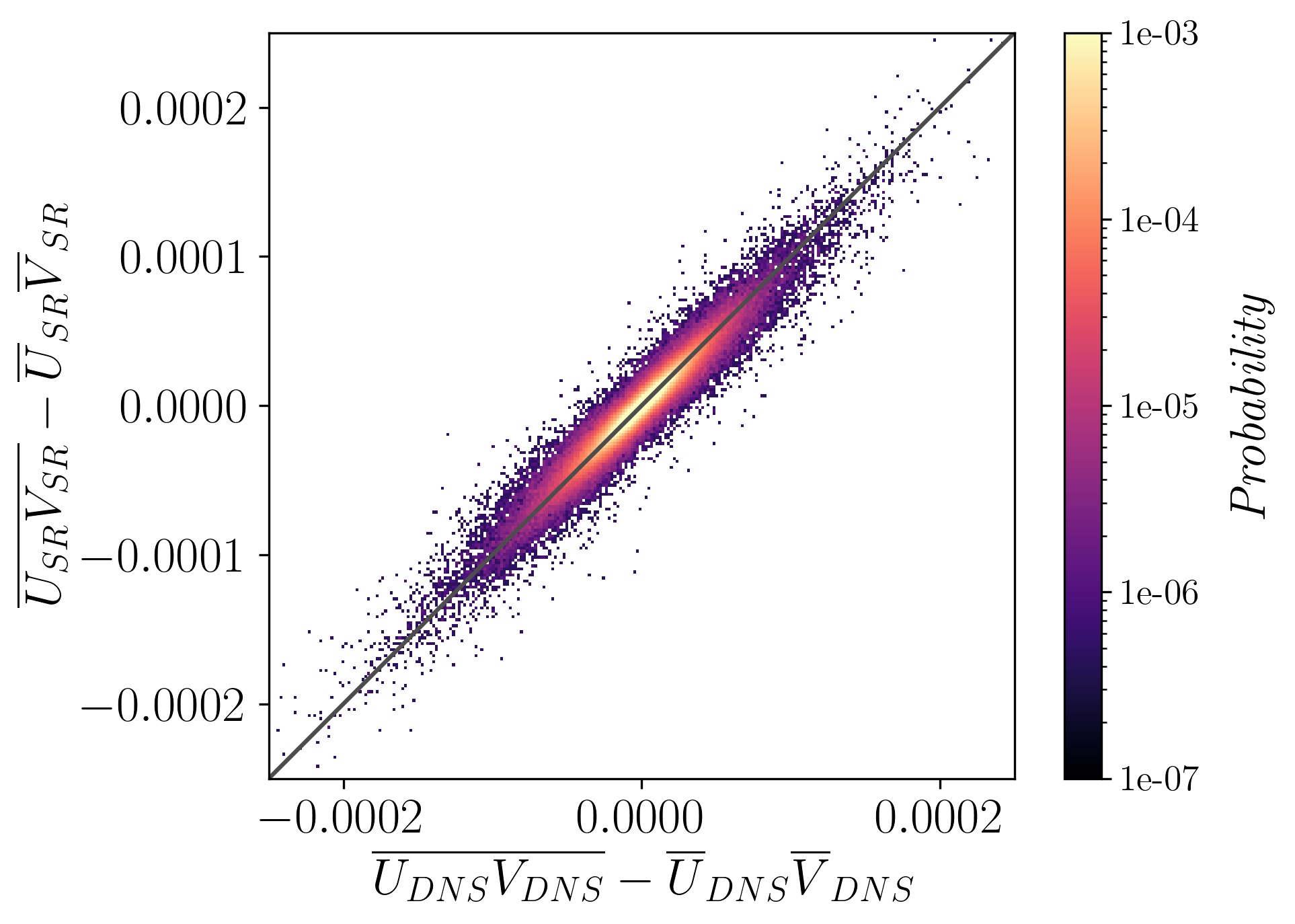}
  
\endminipage
\caption{Joint PDFs of the box-filtered $\tauSFS_{12}$ with PIESRGAN reconstruction when applied to the \textit{Re130} data set (left) and the \textit{Re350} data set (right).}
\label{fig:jPDF_extrapolation}
\end{figure}

The architecture can capture the scale similarity of the turbulent motions and be applied at various Reynolds numbers on similar configurations. It is worth noting that the higher-Reynolds-number reconstruction (\textit{Re350}) is slightly more accurate than the-lower-Reynolds-number reconstruction (\textit{Re130}). This is due to the ratio $\Delta/ \eta$ differing slightly from the one used for training (Sec.~\ref{sec:architectures}). In both extrapolation investigations, the PIESRGAN reconstruction indicates a $17 \%$ reduction in RMSE compared to the reconstruction achieved by TSResNet. The jPDFs of $\tau_{12}^\SFSsub$ evaluated on the  PIESRGAN-reconstructed field for \textit{Re130} and \textit{Re350} inputs are shown in Fig.~\ref{fig:jPDF_extrapolation}. In both cases, the PIESRGAN model extrapolates accurately to higher $Re$, with remarkable alignment of the predicted SFS stress with the DNS data. In both cases, the cross-correlation of every SFS stress tensor component computed from the fields obtained from the PIESRGAN model exceeds $90\%$. 

%%%%%%%%%%%%%%%%% CONCLUSION %%%%%%%%%%%%%%%%%%%%%
\section{Conclusion}

We investigated the influence of adversarial training on super-resolution turbulence reconstruction. GAN-based models (PIESRGANs) are assessed against a standard, supervised CNN-based model (TSResNet). Two approaches are introduced that leverage the GAN model's adversarial training capabilities, enabled by its discriminator network, to enhance the generator network's accuracy and generalizability to out-of-sample inputs for different filter sizes and filter kernels. The GAN generator and TSResNet have identical model structures, hence their evaluation cost is the same despite the enhanced accuracy of the GAN model.
% In CFD we typically associate better models with more computational expense. You can have better, more expensive models, or worse, less expensive ones. Here you introduce a better one, that is not more expensive. That is a major advantage.
% the only difference being the GAN discriminator and the corresponding adversarial contribution in the generator loss function. The GAN generator has an identical structure to the TSResNet. 

The two training approaches have first been evaluated for in-sample data (i.e., statistically similar to the training data). In TKE spectra and velocity-gradient PDFs, the PIESRGAN architecture more accurately reconstructs the full-resolution velocity fields compared to the TSResNet model. The TSResNet model accurately captures the large-scale features, though the PIESRGAN is more consistent throughout the range of scales and velocity increments.  
%In the context of deconvolutional modeling, the TSResNet appears to perform a soft deconvolution operation, reconstructing scales only larger than the LES filter, which would necessitate a SFS model for \textit{a posteriori} calculations. Conversely, the PIESRGAN effectively performs both soft and hard deconvolution simultaneously. Thus a second closure term would not be needed. 

Adversarial training, while computationally more expensive, is promising to enhance SR model reconstruction of small-scale turbulence. In \textit{a priori} assessments of the SFS stress tensor, the SR models outperform the widely used dynamic Smagorinsky model, showing significantly improved statistical alignment with the ``true'' filtered-DNS SFS stresses. The usefulness of these findings will, however, need to be further verified through \textit{a posteriori} tests. Of the SR models, the PIESRGAN-reconstructed SFS fields are 64\,\% better aligned with the filtered-DNS SFS fields than those produced by the TSResNet. This is due entirely to the GAN discriminator's feedback to the generator; all other training was performed identically. The discriminator's semisupervised training enables it to capture more complex, nonlinear relationships between the low- and high-resolution fields than a standard, supervised training approach would permit.

Data-driven super-resolution methods are known to suffer significant performance drops when applied to substantially out-of-sample data. This capability is particularly important to upscale real-world fields. To test this, we investigated two non-consistent applications of the SR models to testing data sets: out-of-sample filter kernels and out-of-sample upscaling ratios. To assess the influence of the training filter kernel, box- and spectrally sharp-filtered SR models were applied to Gaussian-filtered DNS fields. The box/spectral-trained SR models consistently over/underestimated the SFS kinetic energy when applied to Gaussian-filtered fields and produced poorly aligned jPDFs of $\tauSFS_{12}$. To mitigate this behavior,  both architectures were trained for a random collection of fields containing both box- and spectral-filtered data. The ensuing models have slightly greater reconstruction capability than the single-filter-trained models, though both still had a consistent gap to the Gaussian-filtered DNS. The TSResNet was again inaccurate for the subfilter scales, while the PIESRGAN added artifacts at high wave numbers, with slight underprediction of filter-scale features. 

To improve the GAN generator's ability to upsample unlabeled, low-resolution fields (i.e., without accompanying high-resolution fields), computed LES fields were provided as LR training inputs. This choice enabled the generator to refine and augment existing SFS structures only from input fields that possess physical meaning, ensuring an effective contribution to the generative process. The discriminator was employed in an unsupervised manner to provide feedback to the generator during training. By combining inputs with multiple kernels and computed LES data, the generator was trained to produce fields with correct high-frequency details emulating the DNS. This partially unsupervised approach permits the PIESRGAN architecture to handle diverse input fields more effectively, and the issue of consistent small-scale overprediction of energy density is alleviated. 
%To improve the GAN generator's ability to upsample unlabeled low-resolution fields (i.e., without accompanying high-resolution fields), as would occur for \emph{a posteriori} LES, computed LES fields were provided as SR training inputs. The discriminator was employed as a feature extractor to provide feedback to the generator during training. By combining different filter kernels with computed LES data, the generator was trained to produce fields with correct high-frequency details emulating the DNS. This partially unsupervised approach permits the PIESRGAN model to handle diverse input fields better, and the issue of consistent small-scale overprediction is alleviated.

%This is particularly relevant in the context of turbulence modeling, as the model needs to be applicable universally independently of the specific data used for training. In order to assess how SR models perform when a filter kernel different from the one used to obtain training data is used to filter the input data, the pre-trained models trained with fields filtered with a box filter or spectral filter are used to reconstruct a field filtered with a Gaussian kernel. 

The ability of SR models to upsample out-of-sample filter-to-DNS mesh ratios was also explored. When a single upsampling factor is used for training, the reconstructed field may exhibit blurriness and artifacts, leading to inaccurate super-resolved fields and decreased predictive capability. To overcome this issue, two additional upsampling branches were added to the generator, with the discriminator determining which branch to apply locally. This enables the GAN model to manage input fields of potentially unknown resolution. By selecting the upsampling factor that minimizes this adversarial loss, the generator most faithfully reconstructs a given input field, at least within the range of upsampling ratios provided during training. We find that training over different filter kernels, unlabeled data (e.g., true LES data), and multiple filter-to-DNS mesh ratios enables sufficient model robustness and reconstruction capabilities.
%The ability of SR models to upsample out-of-sample filter-to-DNS mesh ratios was also explored. When a single upsampling factor is used for training, the reconstructed field may exhibit blurriness and artifacts, leading to inaccurate super-resolved fields and decreased predictive capability. This is particularly critical for LES closure modeling, where fixed upsampling ratios cannot be presumed in general, particularly on nonuniform meshes. To overcome this issue, two additional upsampling branches were added to the generator, with the discriminator determining which branch to apply locally. This enables the GAN model to manage input fields of potentially unknown resolution. By selecting the upsampling factor that minimizes this adversarial loss, the generator most faithfully reconstructs a given input field, at least within the range of upsampling ratios provided during training. From these analyses, it can be concluded that different filter kernels, unlabeled data (e.g., true LES data), as well as multiple filter-to-DNS mesh ratios must be provided during training to enable sufficient model robustness.

Finally, the ability of the two models to extrapolate to higher Reynolds numbers was tested. Using input fields from higher-Reynolds-number DNS, filtered to maintain the same $\olDelta/\eta$ ratio used for training, the PIESRGAN architecture more consistently reconstructed small-scale flow features than the TSResNet. This suggests that GAN-based models trained for lower Reynolds numbers can be successfully applied to slightly higher Reynolds numbers, though further testing for non-HIT flows (e.g., turbulent shear flows, turbulent reacting flows) is ongoing.

%This approach allows for more flexibility in data-driven models, which may no longer need to be as strongly restricted to relying on DNS data or highly-resolved LES for physical conditions. 

Adversarial training, although computationally more costly than standard supervised approaches, is overall successful in improving the reconstruction and extrapolative capabilities of the data-driven SR model. Leveraging the discriminator's feedback, a GAN-based model is able to generate more diverse and realistic samples during training that improve the architecture's performance and generalizability to unseen training data. %We expect the robustness imparted by adversarial training to be effective in reducing \emph{a posteriori} SFS modeling errors and improving the stability of SR-LES calculations, which will be essential in developing SR data-driven models for SFS closures.
%%How are the examples/samples used to improve the model's performance? Isn't the realistic sample a direct consequence of the model's improved performance?
%and enable a posteriori verification of the generated turbulence fields.
%What does the discriminiator have to do with a-posteriori modelling? Is this not possible with CNNs?

\section*{Acknowledgments}
The research leading to these results has received funding from the European Union’s Horizon 2020 research and innovation program under the Center of Excellence in Combustion (CoEC) project, Grant Agreement No.~\textit{952181}, and from the German Federal Ministry of Education and Research (BMBF) and the state of North Rhine-Westphalia as part of the NHR funding. The authors gratefully acknowledge the computing resources from the DEEP-EST project, which received funding from the European Union’s Horizon 2020 research and innovation program under Grant Agreement no.~\textit{754304} and the computing time granted by the NHR4CES Resource Allocation Board on the high-performance computer CLAIX at the NHR Center RWTH Aachen University (project \textit{rwth0733}). We thank Dr.~R.~Sedona and Dr.~G.~Cavallaro for their support in the porting of the application to DEEP-EST. The authors thank S.~Sakhare, F.~Fr{\"o}de, and Dr.~P.~Petkov for their exceptional support and contributions to this research project.

%%%%%%%%%%%%%%%%%%%%%%%%%%%%%%%%%%%%%%%% APPENDIX %%%%%%%%%%%%%%%%%%%%%%%%%%%%%%%%%%%%%%%

\appendix
\section{Influence of numerical precision}
\label{app:appendix}

``Mixed-precision'' training has emerged as a powerful technique with significant implications for deep learning models. In conventional training, using 32-bit floating-point precision (``single precision,'' \textit{FP32}) for storing model parameters incurs higher memory requirements. To address this limitation and accelerate model training, reduce memory usage, and improve inference times, mixed-precision training has gained popularity. It combines \textit{FP16} (16-bit floating-point precision) and \textit{FP32}, for different parts of the neural network computation. Its major drawback is that reduced numerical precision can considerably reduce the precision of the resulting model, which is particularly detrimental for highly multiscale problems~\cite{Hrycej_2022}.
Lower precision can compromise a trained model's parameters and increase the probability that the magnitude of the loss gradients could be in the same order as the floating-point roundoff error.

\begin{figure}[!ht]
    \centering
    \includegraphics[scale=0.5]{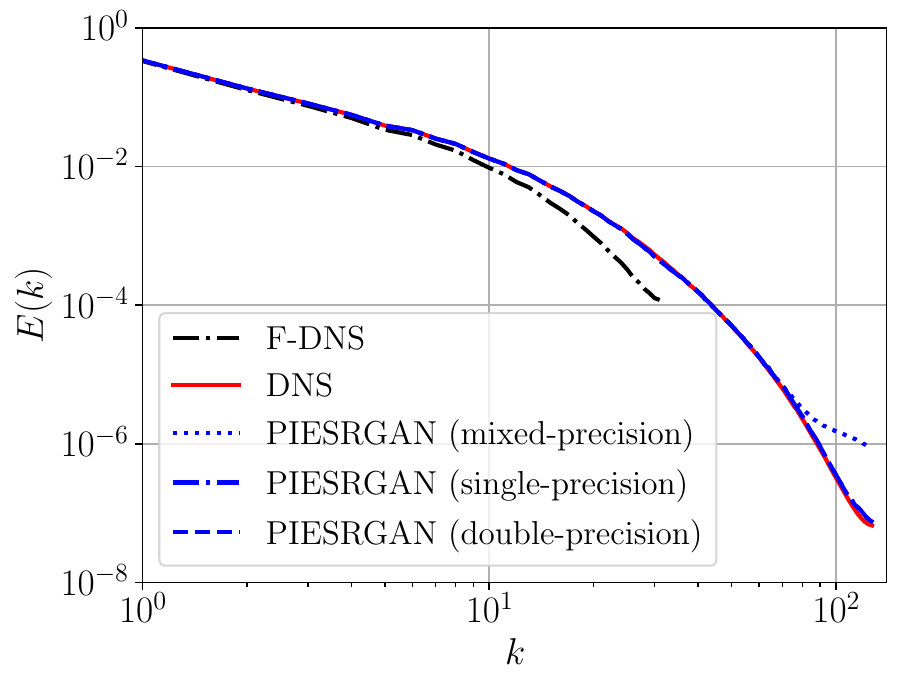}
    \caption{Comparison of the TKE spectra for in-sample box-filtered training/testing fields when different numerical-precision training are adopted. Mixed-precision combines half-precision floating point (\textit{FP16}) with single-precision floating point (\textit{FP32}).}
    \label{fig:FP16vsFP32}
\end{figure}

To evaluate the influence of numerical precision, Fig.~\ref{fig:FP16vsFP32} shows TKE spectra for in-sample, box-filtered, super-resolved fields from PIESRGAN models trained using mixed, single, and double precision (\textit{FP64}).
Mixed-precision training clearly leads to an accumulation of TKE at high wave numbers, while both single and double precision are visually comparable to the DNS at these high wave numbers. Because it is computationally less intensive, single precision is therefore chosen for the model-training results presented throughout the paper.

\bibliographystyle{IEEEtran}  
%\bibliography{jfm-instructions}
\bibliography{sample}

\end{document}